\documentclass[11pt,a4paper]{article}
\pdfoutput=1

\usepackage{jheppub}

\usepackage{amsmath}
\usepackage{bbm}
\usepackage{float}
\usepackage{graphicx}	
\usepackage{hyperref}
\usepackage{mathtools}
\usepackage{cancel}

\usepackage[normalem]{ulem}

\usepackage{multirow}
\usepackage{rotating}
\usepackage{subcaption}

\def\lsim{\raise0.3ex\hbox{$\;<$\kern-0.75em\raise-1.1ex
\hbox{$\sim\;$}}}
\def\gsim{\raise0.3ex\hbox{$\;>$\kern-0.75em\raise-1.1ex
\hbox{$\sim\;$}}}

\def\thetitle{
A framework for testing leptonic unitarity by neutrino
oscillation experiments 
 {\small \vskip -4.4cm \hglue 11cm \rm YACHAY-PUB-16-02-PN}
 \vspace{2.5cm}
}
\title{\thetitle}
\hypersetup{pdftitle={\thetitle}}

\author{Chee Sheng Fong$^{1}$}
\author{Hisakazu Minakata$^{2}$}
\author{Hiroshi Nunokawa$^{3}$}
\affiliation{
$^1$Instituto de F\'{\i}sica, Universidade de S\~ao Paulo, C.\ P.\
66.318, 05315-970 S\~ao Paulo, Brazil \\
$^2$Department of Physics, 
Yachay Tech, San Miguel de Urcuqu\'i, 100119 Ecuador  \\
$^3$Departamento de F\'{\i}sica, Pontif{\'\i}cia Universidade Cat{\'o}lica 
do Rio de Janeiro, C. P. 38097, 22451-900, Rio de Janeiro, Brazil  \\
}
\emailAdd{fong@if.usp.br}
\emailAdd{hminakata@yachaytech.edu.ec}
\emailAdd{nunokawa@puc-rio.br}

\abstract{ 
If leptonic unitarity is violated by new physics at an energy scale much
lower than the electroweak scale, which we call low-scale unitarity
violation, it has different characteristic features from those expected
in unitarity violation at high-energy scales. They include maintaining
flavor universality and absence of zero-distance flavor transition. We
present a framework for testing such unitarity violation at low energies
by neutrino oscillation experiments. Starting from the unitary 3 active plus
$N$ (arbitrary positive integer) sterile neutrino model we show that 
by restricting the active-sterile and sterile-sterile neutrino mass squared differences to $\gsim$ 0.1 eV$^2$ 
the oscillation probability in the $(3+N)$ model
becomes insensitive to details of the sterile sector, providing a nearly
model-independent framework for testing low-scale unitarity
violation. Yet, the presence of the sterile sector leaves trace as a
constant probability leaking term, which distinguishes low-scale
unitarity violation from the high-scale one. The non-unitary mixing
matrix in the active neutrino subspace is common for the both cases. We
analyze how severely the unitarity violation can be constrained in
$\nu_{e}$-row by taking a JUNO-like setting to simulate medium baseline
reactor experiments. Possible modification of the features of the
$(3+N)$ model due to matter effect is discussed to first order in the matter potential. 
}

\DeclareMathOperator{\diag}{diag}

\begin{document} 

\maketitle

\section{Introduction}
\label{sec:introduction}

Determination of leptonic mixing parameters, the three mixing angles and the two mass squared differences of neutrinos, marks a new epoch of physics beyond the standard model (SM). Despite that we still do not know the value of leptonic Kobayashi-Maskawa phase and the neutrino mass pattern, successes of {\em hypothesis} of using the standard three-flavor mixing in describing a wealth of experimental data prompts us to think about one further step, namely, the paradigm test. We feel that we have reached the right point that raising the question of how to test the three-flavor mixing framework itself is timely. 

The most common way of testing the framework is to verify unitarity of the mixing matrix. It appears to us that the two different strategies of testing leptonic unitarity are thinkable: 
\begin{itemize}
\item
One is to show closing of the lepton unitarity triangle in an analogous way of unitarity test for the quark CKM matrix \cite{Agashe:2014kda}.  

\item
The other is to prepare a model of unitarity violation, confront it against the available experimental data, and derive constraints on unitarity violation parameters.

\end{itemize}
\noindent 
The first method serves as a unitarity test purely within the framework of standard three-flavor mixing scheme of neutrinos, without recourse to any particular model of unitarity violation. This is the advantage of this method. See e.g., ref.~\cite{Farzan:2002ct} for this approach. In the second way, one introduces a general framework for leptonic unitarity violation, or a class of models that embody the property is constructed. 

To our opinion, there are pros and cons in the above two different strategies. In the first method, despite its charming model-independent nature, it is quite challenging to determine size of each side of the unitarity triangles experimentally \cite{Farzan:2002ct}. The second method introduces model-dependent features into the unitarity test, which can be considered as a drawback of this approach. On the other hand, there is a definite underlying scenario behind the non-unitarity in the latter case, or at least general guidance to it. Therefore, once hinted, it may allow us to identify the cause of unitarity violation. We feel, therefore, that both methods of leptonic unitarity test must be pursued.

In studies of testing leptonic unitarity so far done along the second strategy above, it appears that people took two different attitudes.
That is, one integrates out new physics effects at high energy scales to obtain effective theories of three generation leptons at low energies which represents non-unitarity in this limited subspace \cite{Antusch:2006vwa}. The other one takes more relaxed attitude in explicitly introducing SM singlet leptons, and examines the models in a relatively model independent fashion in SM subspace with non-unitarity \cite{Escrihuela:2015wra,Parke:2015goa}. In the latter approach, the masses of SM gauge-group singlet leptons can be large or small, reflecting varying underlying scenarios of new physics. If we take the former way, $SU(2) \times U(1)$ gauge invariance dictates that the same unitarity violation must also be manifest in the charged lepton sector. In the framework of ref.~\cite{Antusch:2006vwa}, generally speaking, the constraints on unitarity violation are dominated by the ones coming from the charged lepton sector. If we follow the latter way, it is more case-sensitive and neutrino experiments can play greater roles. One of the most interesting questions in the whole area of study of unitarity violation is to reveal qualitative differences between unitarity violation at high- and low-energy scales. 

It is the purpose of this paper to discuss low energy-scale unitarity violation, hereafter {\em ``low-scale unitarity violation''} for short, in detail. By low-scale unitarity violation we mean that a ``hidden'' sector in state space to which probability flow occurs is located at low energy scales, like eV or MeV. It allows the hidden sector particles be produced along with neutrinos, and also they participate in neutrino oscillations assuming their mixing with neutrinos. We first recapitulate the interesting characteristic features of low-scale unitarity violation different from high-scale unitarity violation. They include (1) retaining flavor universality, and (2) lack of zero-distance flavor transitions. See section~\ref{sec:unitarity-high-low} for more about these points. 
Some specific scenarios of high- and low-scale unitarity violation were explored in refs.~\cite{Xing:2012kh,Luo:2014fia,Luo:2014vha,Li:2015oal}.

Then, we go on to construct a framework for experimental testing of low-scale unitarity violation. Since there is such interesting qualitative differences above between high- and low-scale unitarity violation, they must be tested and be distinguished from each other.
We argue, in agreement with the preceding works \cite{Parke:2015goa,Escrihuela:2015wra}, that the constraint by $Z$ width measurement in LEP \cite{LEP} makes extension of low-mass lepton sector to be essentially unique, only allowing inclusion of SM singlet fermions. 
Thus, our model of non-unitarity at low-energies utilizes three active neutrinos and an arbitrary numbers of sterile neutrino states.

We discuss in detail how the model prediction can be made insensitive to the details of the sterile sector, e.g., the mass spectrum of sterile neutrinos and mixing between active and sterile neutrinos.
We find that the resultant expressions of oscillation probabilities in
vacuum contain a new term, an explicit probability leakage term, which
distinguishes between low- and high-scale unitarity violation. To our
knowledge, the term has not been incorporated in the previous analyses
of unitarity violation at low energies. We examine how this framework
works by analyzing future medium-baseline reactor neutrino experiments. 
In the final two sections we discuss how CP violating terms in
accelerator appearance measurement can be used to signal
non-unitarity and how the matter effect affects the foregoing discussions above. 

There is an obvious relation between the model we discuss in this paper and various versions of active plus sterile neutrino models proposed to provide description of the LSND-MiniBooNE anomaly (see a review \cite{Conrad:2013mka} and references therein). We will make remarks on the relationship between them below at wherever appropriate. In particular, we should note that in the frameworks of 3 active plus a few sterile neutrinos the various bounds on the mixing parameters are derived by using the existing data.  For the most recent comprehensive analysis, see ref.~\cite{Kopp:2013vaa}.\footnote{
Though they are very relevant for this paper we do not implement these bounds into discussions in this paper. It is because they are not derived by using the generic $(3+N)$ model, and the translation of their bound to our setting requires great care. Furthermore, the principal purpose of this paper is to provide suitable framework for leptonic unitarity test in high-precision experiments in the future.}

\section{Unitarity violation at high- and low-energy scales}
\label{sec:unitarity-high-low}

The cause of unitarity violation in the lepton sector can be due to new 
physics beyond the SM at high-energy scales, or the ones at low energies. In the best studied high-scale seesaw scenario of neutrino mass \cite{Minkowski:1977sc,Yanagida:1979as,GellMann:1980vs,Mohapatra:1980yp}, the three-flavor mixing of neutrinos has a tiny violation of unitarity due to the mixing of heavy right-handed neutrinos. A more generic formulation of high-scale unitarity violation was given by Antusch et al.~\cite{Antusch:2006vwa} by taking the minimal unitarity violation scheme. One of the salient features in high-scale unitarity violation is that even though SM singlet leptons exist which mix with neutrinos, it is likely that such neutral leptons are much heavier than neutrinos. They are not produced copiously in the same processes as neutrinos are produced, and a physical transition from neutrinos to them are kinematically forbidden.

On the other hand, if we assume that unitarity violation occurs due to physics at an energy scale much lower than the electroweak scale, the light SM gauge group singlet leptons not only mix with neutrinos, but also their masses are so light that they participate in the process of neutrino oscillations. In this paper, we try to develop a framework for experimental test of unitarity violation by assuming such situation, to which we simply refer as ``low-scale unitarity violation''. Hereafter, we call the SM gauge group singlet fermions generically as ``sterile neutrinos'' for simplicity.

We notice that there are some characteristic features in high- and low-scale unitarity violation that one can recognize even without going into any details. They are:

\begin{itemize}

\item
Yes or no of violation of lepton universality:  
It is expected on general ground that due to non-unitarity of the lepton mixing matrix the lepton universality is violated. See refs.~\cite{Antusch:2006vwa,Escrihuela:2015wra}, and the references cited therein.\footnote{
General bounds on non-unitarity are discussed in the context of high-scale unitarity violation, e.g., in refs.~\cite{Antusch:2006vwa,Escrihuela:2015wra,Antusch:2014woa,Fernandez-Martinez:2016lgt}. 
}
While it is a generic feature in high-scale unitarity violation, lepton universality can be maintained in low-scale unitarity violation. It is because sterile neutrinos can be produced as well, for example in $\mu \rightarrow e + \text{steriles}$ process. Assuming no detection sensitivity to sterile neutrinos, it masks the effect of non-unitary mixing matrix in the active neutrino sector.  

\item
Yes or no of zero distance neutrino flavor transition: 
Similarly to the above point, in high-scale unitarity violation, kinematically forbidden active to sterile states transition entails zero-distance attenuation of probability of a given flavor neutrino \cite{Langacker:1988ur}. It {\em does not} occur if sterile neutrinos can take part in the flavor oscillation processes, as we will show in section~\ref{sec:zero-distance}.

\item
Of course, there are common features in high- and low-scale unitarity violation: Emission of sterile neutrinos, if kinematically allowed range of low to high masses, affects the observable spectrum of charged leptons. It can be utilized to place constraints on non-unitarity by using, e.g., electron spectra in beta and muon decays, or muon spectrum in pion decay, cosmological observations etc. See \cite{deGouvea:2015euy} for a comprehensive summary of the current status of the bounds for the 3+1 scenario.\footnote{
For some of the early analyses of extra neutral heavy leptons and the bounds on them, see e.g., \cite{Langacker:1988ur,Nardi:1991rg,Nardi:1994iv}.  }
\end{itemize}
\noindent 
In the rest of this paper, we construct a model of low-scale unitarity violation which can be used to test leptonic unitarity in neutrino experiments. Although the constraints from beta and muon decays etc. just mentioned above are relevant, we do not try to elaborate the discussions already given in \cite{deGouvea:2015euy} and the references cited therein.

\section{A model of unitarity violation at low energies}
\label{sec:low-E-unitarity-violation}

Now, we introduce our model of unitarity violation at low energies. 
But, one recognizes immediately that there is no big room for this. Precision measurement of $Z$ decay width at LEP \cite{LEP} dictates that there is only three active neutrinos. Therefore, extra fermions we introduce which mix with neutrinos must be SM singlets, which we call ``sterile neutrinos'' in this paper. Then, we are left with the unique possibility, the system of three active neutrinos plus arbitrary number of sterile neutrinos which mix with each other. We denote the number of sterile neutrino states as $N$. We assume that our system of the three active neutrinos and $N$ sterile neutrinos are complete, which we call the {\bf $(3+N)$ space unitary model} hereafter. 

Though we deal with the particular model we want it as {\em model-independent} as possible within the framework of the $(3+N)$ space unitary model. Therefore, we shall always keep number of sterile neutrinos $N$ arbitrary in this paper. Toward constructing a framework for leptonic unitarity test, however, we must make additional requirement on our $(3+N)$ space unitary model. We want to avoid the situation that experimental predictions of the model depend very sensitively on details of the $N$ sterile neutrino sector, for example, on the mass spectrum of sterile states. In the rest of this section, we discuss how it can be achieved, and what are the conditions for this. 

\subsection{3 active $+ N$ sterile unitary system}
\label{sec:non-unitarity-vacuum}

To define the notation and for definiteness, we introduce the $(3+N)$
space unitary system in vacuum. The Hamiltonian which governs the
evolution of 3 active and $N$ sterile state vector in flavor basis, 
$\nu = \left[ \nu_{e}, \nu_{\mu}, \nu_{\tau}, \nu_{s_{1}}, \nu_{s_{2}},
\cdot \cdot \cdot, \nu_{s_{N}} \right]^{T}$, 
as $ i \frac{d}{dx} \nu = H \nu$ is given by\footnote{
For simplicity, we assume that sterile states do not decay along its length of flight in neutrino oscillation experiments. If sterile states have decay length much shorter than the baseline, and if the decay products do not include the three active neutrinos, the oscillation probabilities converge to those of ``high-scale unitarity violation'' discussed in the previous section. 
} 
\begin{eqnarray}
H = {\bf U} \left[
\begin{array}{cccccc}
\Delta_{1} & 0 & 0 & 0 & 0 & 0 \\
0 & \Delta_{2} & 0 & 0 & 0 & 0 \\
0 & 0 & \Delta_{3} & 0 & 0 & 0 \\
0 & 0 & 0 & \Delta_{4} & 0 & 0 \\
0 & 0 & 0 & 0 & \cdot \cdot \cdot & 0 \\
0 & 0 & 0 & 0 & 0 & \Delta_{3+N} \\
\end{array}
\right] {\bf U}^{\dagger} 
\label{H-3+N-vac}
\end{eqnarray}
where 
\begin{eqnarray}
\Delta_{i}  \equiv \frac{ m^2_{i} }{2E} 
\hspace{4mm}
(i = 1,2,3),
\hspace{8mm}
\Delta_{J}  \equiv \frac{ m^2_{J} }{2E}
\hspace{4mm}
(J = 4, \cdot \cdot \cdot, 3+N).
\label{Delta-def}
\end{eqnarray}
Here, $m_{i}$ ($m_{J}$)
denote the mass of mostly active (sterile) neutrinos and $E$ is the neutrino energy. 

For notations of the mass squared differences we use, in generic case including both active and sterile neutrinos,  
\begin{eqnarray}
\Delta m_{ab}^2 \equiv m_a^2 - m_b^2,
\label{eq:mass_diff-generic}
\end{eqnarray}
where $a,b=1,...,3+N$. When we want to distinguish between active-active,  active-sterile, and sterile-sterile neutrino mass differences, we use 
\begin{eqnarray}
\Delta m_{ji}^2 &\equiv& m_j^2 - m_i^2, 
\nonumber \\ 
\Delta m_{J i}^2 &\equiv& m^2_{J} - m^2_{i}, 
\nonumber \\ 
\Delta m_{J I}^2 & \equiv & m^2_{J} - m^2_{I}, 
\label{eq:mass_diff-specific}
\end{eqnarray}
where $i,j,k, ...$ (small letters) = 1,2,3 are for active neutrinos, 
and $I,J,K, ... $ (capital letters) = 4,.., 3+$N$ 
for sterile neutrinos.

The mixing matrix ${\bf U}$ relates the flavor eigenstate $\nu$ to the
vacuum mass eigenstate $\tilde{\nu}$ as 
$\nu_{\zeta} = {\bf U}_{\zeta a} \tilde{\nu}_{a}$ (here, $\zeta=e,\mu,\tau,s_1,...,s_N$), 
and hence it is a $(3+N) \times (3+N)$ matrix. 
By construction of the model, the matrix {\bf U} is unitary. 
While the flavor index $\zeta$ above includes also sterile states, 
from now on, we will eventually single out only the active 
flavor indices $\alpha, \beta = e, \mu, \tau$ whenever they are 
explicitly specified in the formulas for the $S$ matrix as well as for
probabilities.

\subsection{Averaging out the sterile oscillations due to decoherence} 
\label{sec:decoherence}

How to make our model insensitive to the mass spectrum in the sterile
sector? 
The masses of the sterile neutrinos appear in the factor of 
$e^{-i E_{J} x} = e^{-i \sqrt{p_{J}^2 + m_{J}^2} x}$
in the propagations of mass eigenstates. This results in phase differences 
$e^{-i (E_{J} - E_{I}) x} \approx e^{-i \Delta m_{JI}^2 x/(2E)}$ 
which can be observed through neutrino oscillation phenomena. 
Assuming no accidental mass degeneracy among the sterile states i.e. 
$|\Delta m_{JI}^2| \gg |\Delta m_{31}^2|$, the oscillation terms involving sterile masses can be averaged out due to (partial) decoherence if certain conditions are fulfilled.\footnote{
We are interested in partial decoherence where the oscillations involving sterile states are averaged out, whereas the active ones do oscillate. 
}
Intuitively, decoherence occurs when the variation in the phase due to
spatial and/or energy resolution is greater than $2\pi$, 
\begin{eqnarray}
\left| \delta \left(\frac{\Delta m_{ab}^2 x}{2E}\right) \right| 
= \left| \frac{\Delta m_{ab}^2}{2E} \delta x 
- \frac{ \Delta m_{ab}^2 x}{2E^2} \delta E \right| \gtrsim 2\pi.
\label{eq:decoherence}
\end{eqnarray}
From the terms in eq.~\eqref{eq:decoherence} which depend, respectively, on the variation of baseline distance ($\delta x$) and that of energy ($\delta E$), we can classify the following two types of decoherence:
\begin{enumerate}
\item[i.] \emph{Spatial resolution.} In this case, decoherence happens if
\begin{eqnarray}
\delta x \gtrsim \frac{4 \pi E}{|\Delta m_{ab}^2|}.
\label{eq:decoherence_x}
\end{eqnarray}

\item[ii.] \emph{Energy resolution.} In this case, decoherence happens if
\begin{eqnarray}
\delta E \gtrsim \frac{4 \pi E^2}{|\Delta m_{ab}^2| x}.
\label{eq:decoherence_E}
\end{eqnarray}

\end{enumerate}
Notice that the conditions derived heuristically above are in agreement with
those obtained from formal approaches (i.e. wavepacket description) as
e.g., in refs.~\cite{Giunti:1997wq,Hernandez:2011rs,Akhmedov:2012uu}. Since we
are interested in the decoherence involving sterile sector, the
conditions \eqref{eq:decoherence_x} and \eqref{eq:decoherence_E} have to
be fulfilled for $\Delta m_{Ja}^2$ which involve at least one sterile
mass. This allows us to obtain \emph{lower} bound on the scale of
sterile sector $|\Delta m_{Ja}^2|$ where our model becomes insensitive
to the sterile mass spectrum. Notice that $\delta x$ and $\delta E$ in
eqs.~\eqref{eq:decoherence_x} and \eqref{eq:decoherence_E} are
associated with the experimental setup (concerning \emph{both}
production and detection), 
e.g., with the production region of neutrinos and with energy resolution of a detector.

In the following, we discuss the conditions that must be satisfied for
sterile oscillations to be averaged out. Most of the neutrino
oscillation experiments work with the kinematical setting, either 
\begin{eqnarray}
\frac{\Delta m^2_{21} x}{4E} \sim 1
\hspace{10mm}
\text{or}
\hspace{10mm}
\frac{|\Delta m^2_{31}| x}{4E} \sim 1. 
\hspace{10mm}
\label{VOM}
\end{eqnarray}
in which the former is for long-baseline (LBL) reactor neutrino experiments, KamLAND, JUNO, and RENO-50, and the latter for the accelerator LBL and the reactor $\theta_{13}$ experiments. 

From eq.~\eqref{eq:decoherence_x}, the condition for averaging out due
to the size of production region, i.e. $\delta x = x_{\rm prod}$ reads
\begin{eqnarray}
x_{\rm prod} \gtrsim \frac{4 \pi E}{ |\Delta m_{Ja}^2| },
\label{prod-region-average}
\end{eqnarray}
where 
$x_{\rm prod}$ denotes the size of the production region of neutrinos, e.g., core diameter for nuclear reactor neutrinos and the length of decay pipe for accelerator neutrino beams. Assuming the setting as in (\ref{VOM}), the condition (\ref{prod-region-average}) leads to 
\begin{eqnarray} 
|\Delta m^2_{Ja}| &\gsim& 
\pi \Delta m^2_{21} 
\left( \frac{ x }{ x_{\rm prod} } \right) 
\approx 
1.2~\text{eV}^2 
\left( \frac{ x / x_{\rm prod} }{ 5 \times 10^{3} } \right) 
\hspace{6mm} \text{(for reactor)},
\nonumber \\
|\Delta m^2_{Ja}| &\gsim& 
\pi |\Delta m^2_{31}| 
\left( \frac{ x }{ x_{\rm prod} } \right) 
\approx 
7.5~\text{eV}^2 
\left( \frac{ x / x_{\rm prod} }{ 10^{3} } \right) 
\hspace{6mm} \text{(for accelerator)},
\label{prod-size}
\end{eqnarray}
where we have taken, for the typical source sizes and baseline
distances, $x_{\rm prod} = 10$ m and $x=50$ km for reactor neutrinos, 
and $x_{\rm prod} = 1$ km and $x=1000$ km for accelerator neutrinos.

Due to energy resolution of a detector, using $\Delta m^2_{21} = 7.5 \times 10^{-5}$ eV$^2$, and $|\Delta m^2_{31}| = 2.4 \times 10^{-3}$ eV$^2$, eqs.~\eqref{eq:decoherence_E} and (\ref{VOM}) lead to the condition on $\Delta m^2_{J a}$ for the sterile oscillation to be averaged out: 
\begin{eqnarray} 
|\Delta m^2_{J a}| &\gsim& \pi 
\left( \frac{ \Delta m^2_{21} }{ \delta E / E } \right) 
\approx 
7.9 \times 10^{-3} \text{eV}^2 
\left( \frac{ \delta E / E }{ 0.03 } \right)^{-1}
\hspace{6mm} \text{(for LBL reactor)},
\nonumber \\
|\Delta m^2_{J a}| &\gsim& \pi 
\left( \frac{|\Delta m^2_{31}|}{ \delta E / E } \right) 
\approx 
7.5 \times 10^{-2} \text{eV}^2 
\left( \frac{ \delta E / E }{ 0.1 } \right)^{-1}
\hspace{6mm} \text{(for accelerator)}.
\label{E-resolution2}
\end{eqnarray}
The aggressive choice of a typical 3\% error in energy measurement in
the former case is based on the JUNO proposal in \cite{JUNO}, whereas a
conservative choice of $\delta E / E = 10\%$ is made for accelerator
neutrino experiments. Therefore, if we restrict ourselves to $\Delta
m^2_{J i} \sim |\Delta m^2_{JK}| \gsim 0.1~\text{eV}^2$, the fast
oscillation due to the active-sterile and sterile-sterile mass squared 
differences can be averaged out by the effect of energy resolution. 
For the JUNO-like setting the requirement on $|\Delta m_{Ja}^2|$ can be 
relaxed by an order of magnitude.

We note that effect of averaging over the production points of neutrinos 
is less sizeable compared to that of energy resolution for the fast sterile 
oscillation to be averaged out, which therefore leads to more restrictive 
condition on the sterile state masses.

\subsection{Requirement on the sterile mass spectrum}
\label{sec:requirement}

In addition to the condition of averaging out the fast sterile
oscillations, we require that the masses of sterile neutrinos are light
enough such that they can be produced in the same environment as
neutrinos are produced. We do this because it is the most significant
characteristic feature of unitarity violation at low energies. It gives
raise to the condition 
$m_{J} \lsim 1$
MeV for reactor neutrinos, and 
$m_{J} \lsim 100$ 
MeV for accelerator neutrinos. 

To summarize: In seeking the case that neutrino oscillation in our 3 active + $N$ sterile neutrino system is insensitive to the detailed properties of the sterile sector, such as the mass spectrum of the sterile states and the fine structure of the active-sterile mixing, we require for sterile neutrino masses in our $(3+N)$ space unitary model that 
\begin{eqnarray}
0.1~\text{eV}^2 \lsim  m_{J}^2 \lsim 1~\text{MeV}^2. 
\label{sterile-mass}
\end{eqnarray}
The lower limit is from condition of averaging out the fast oscillations for accelerator neutrinos, and the upper one from producibility of sterile neutrinos in reactors. 

With the conditions \eqref{eq:decoherence_x} and/or \eqref{eq:decoherence_E} of averaging out the fast oscillations being satisfied we can make approximation\footnote{
To describe the borderline regime, one has to resort to formal description e.g. of refs.~\cite{Giunti:1997wq,Hernandez:2011rs,Akhmedov:2012uu}.
}
\begin{eqnarray}
\left\langle \sin \left(\frac{ \Delta m^2_{J i} x}{ 2E }\right) \right\rangle &\approx& 
\left\langle \sin \left(\frac{ \Delta m^2_{J K} x}{ 2E }\right) \right\rangle \approx 0, \label{average-out1} \\
\left\langle \cos \left(\frac{ \Delta m^2_{J i} x}{ 2E }\right) \right\rangle &\approx& 
\left\langle \cos \left(\frac{ \Delta m^2_{J K} x}{ 2E }\right) \right\rangle \approx 0,
\label{average-out2}
\end{eqnarray}
where $\langle ... \rangle$ stands for averaging over neutrino energy 
within the uncertainty of energy resolution, as well as averaging over uncertainty of distance between production and detection points of neutrinos.  
The latter approximate equalities in \eqref{average-out1} and \eqref{average-out2} assume that there is no accidental degeneracy among the  sterile state masses i.e. $|\Delta m^2_{JK}| \gg |\Delta m^2_{31}|$.

\subsection{Cases in which sterile oscillations are not averaged out}

While allowing wide range of sterile lepton masses, the condition (\ref{sterile-mass}) excludes the certain characteristic regions of sterile neutrino mass spectrum. Exclusion of higher masses is done under the spirit of low-scale unitarity violation, and therefore we consider it granted. But, there is no a priori reason for excluding sterile neutrino masses in regions $\Delta m^2_{J i} \sim |\Delta m^2_{J K}| \sim $ the atmospheric $\Delta m^2$, or the solar $\Delta m^2$. In this case, however, one must expect severe model dependence in the experimental predictions by the $(3+N)$ unitary model. Clearly, the number of CP violating phases depends on $N$, and the additional phases will play important roles in fitting data. Therefore, an extensive separate treatment is necessary to include this case.

How about sterile neutrino masses which are much lighter than the
atmospheric, or the solar $\Delta m^2$? Again, there is no a priori
model-independent reason for excluding this case. If the active neutrino
masses are such that KATRIN can detect the signal, 
$m_{i} \gsim 0.2$ eV, then, averaging out condition may be barely maintained for the fast active-sterile oscillation for very small sterile masses, $\Delta m^2 \simeq 0.04$ eV$^2$. However, it would be accompanied by extremely slow developing (as a function of propagation distance $x$) sterile-sterile oscillations. Clearly, a separate analysis is needed to know to what extent the case survives a test with the currently available experimental data. 

As a summary of our discussions in this section, we state as follows: If
we construct $(3+N)$ space unitary system as a model of low-scale
unitarity violation, we can make the model predictions insensitive to
details of the sterile neutrino sector, such as the mass spectrum. It
requires us to restrict ourselves to the region of sterile neutrino masses 
$0.1~\text{eV}^2 \lsim m_{J}^2 \lsim 1~\text{MeV}^2$.
We assume this in our all subsequent discussions in this paper.

\section{The oscillation probabilities in 3 active and $N$ sterile model in vacuum}
\label{sec:oscillation-P-vac}

Given the Hamiltonian in (\ref{H-3+N-vac}), it is straightforward to compute the neutrino oscillation probabilities $P(\nu_\beta \rightarrow \nu_\alpha)$ in vacuum, where the Greek indices $\alpha, \beta, ... = e, \mu, \tau$. Let us start by showing that there is no zero distance transition in our $(3+N)$ space unitary model. 

\subsection{No zero distance transition in $(3+N) \times (3+N)$ unitary system}
\label{sec:zero-distance}

The oscillation probabilities take the form 
\begin{eqnarray}
P(\nu_\beta \rightarrow \nu_\alpha) &=  & 
\left| ~\sum_a {\bf U}_{\alpha a} {\bf U}^*_{\beta a}  ~e^{-i\frac{m_a^2
 x}{2E}} ~\right|^2 \nonumber\\
 &=  & 
\left| ~\sum_{a = 1}^{3+N} {\bf U}_{\alpha a} {\bf U}^*_{\beta a} ~\right|^2 
- 2 \sum_{b \neq a}  \mbox{Re}[{\bf U}_{\alpha a} {\bf U}_{\beta a}^*
{\bf U}_{\alpha b}^* {\bf U}_{\beta b}] 
\sin^2 \left(\frac{ \Delta m^2_{ba} x}{ 4E }\right)\nonumber\\
&- &\sum_{b \neq a} 
\mbox{Im}[ {\bf U}_{\alpha a} {\bf U}_{\beta a}^* {\bf U}_{\alpha b}^* {\bf U}_{\beta b} ] 
\sin \left(\frac{ \Delta m^2_{ba} x}{ 2E } \right).
\label{eq:P-ba-general}
\end{eqnarray}
At $x=0$, $P(\nu_\beta \rightarrow \nu_\alpha) = \delta_{\alpha \beta}$ thanks to unitarity of the ${\bf U}$ matrix. It means, of course, no zero distance transition in the $(3+N)$ space unitary model. This is in sharp contrast to the feature possessed by the high-scale unitarity violation \cite{Antusch:2006vwa,Escrihuela:2015wra}.

\subsection{The oscillation probabilities in the ($3 + N$) model}
\label{sec:oscillation-P}

Here, we derive the expressions of the oscillation probabilities in our ($3 + N$) model when the active-sterile and sterile-sterile oscillations are averaged out. For this purpose we define a new notation of the $(3+N) \times (3+N)$ unitary matrix {\bf U}. It can be parameterized as~\cite{Escrihuela:2015wra}
\begin{eqnarray} 
{\bf U} = 
\left[
\begin{array}{cc}
U & W \\
Z & V \\
\end{array}
\right],
\label{U-parametrize}
\end{eqnarray}
satisfying ${\bf U} {\bf U}^\dagger = {\bf U}^\dagger{\bf U} = 1_{(3+N)\times (3+N)}$.
The active space mixing matrix $U$ is $3 \times 3$ matrix with elements
$U_{\alpha i}$, the rectangular matrices $W$ and $Z$ are respectively $3
\times N$ and $N \times 3$ matrices 
with elements $W_{\alpha I}$ and $Z_{I \alpha}$, and the square matrix $V$ is $N \times N$ matrix with elements $V_{I J}$. 
To develop general framework we do not make any assumptions on the size of $W$ and $Z$ matrix elements 
(besides $|W|, |Z| < 1$) in this paper. 

The oscillation probability is written in terms of $S$ matrix as $P(\nu_\beta \rightarrow \nu_\alpha) = \vert S_{\alpha \beta} \vert^2$, 
\begin{eqnarray} 
S_{\alpha \beta} &=& 
\sum_{k=1}^{3} U_{\alpha k} U^{*}_{\beta k} e^{ - i \Delta_{k} x} 
+ \sum_{K=4}^{3+N} W_{\alpha K} W^{*}_{\beta K} e^{ - i \Delta_{K} x}.
\label{S-elements-3+N-vac}
\end{eqnarray}
where $\Delta_{k(K)} \equiv { m^2_{k(K)} }/{(2E)}$ as defined in
(\ref{Delta-def}),
and the elements of $3\times N$ matrix $W$ are defined as 
\begin{eqnarray} 
W \equiv
\left[
\begin{array}{ccc}
W_{e\, 4} &\ W_{e\,5} &...\ \ W_{e\,3+N}\\
W_{\mu\, 4} &\ W_{\mu\, 5} &...\  \ W_{\mu\, 3+N}\\
W_{\tau\, 4} &\ W_{\tau\, 5} &...\ \  W_{\tau\, 3+N}
\end{array}
\right],
\label{W-parametrize}
\end{eqnarray}
such that its integer index indicated by the capital letters like
$I,J$ and $K$ (which run from 4 to $3+N$) 
always refers to the sterile neutrino mass eigenstate.

After squaring the $S$ matrix, $P(\nu_\beta \rightarrow \nu_\alpha)$ has three terms: 
the first and second terms squared and the interference term, 
each of which can be easily computed. They are given, in order, as   
\begin{eqnarray}
&& P(\nu_\beta \rightarrow \nu_\alpha) =  
\left| \sum_{k=1}^{3} U_{\alpha k} U^{*}_{\beta k} \right|^2 
\nonumber\\
&-&
2 \sum_{j \neq k} 
\mbox{Re} 
\left( U_{\alpha j}^* U_{\beta j} U_{\alpha k} U^{*}_{\beta k} \right) 
\sin^2 \frac{ ( \Delta_{k} - \Delta_{j} ) x  }{ 2 } 
+ \sum_{j \neq k} \mbox{Im} 
\left( U_{\alpha j}^* U_{\beta j} U_{\alpha k} U^{*}_{\beta k} \right) 
\sin ( \Delta_{k} - \Delta_{j} ) x 
\nonumber\\
&+& 
\sum_{J} 
\vert W_{\alpha J} \vert^2 \vert W_{\beta J} \vert^2 
\nonumber\\
&+&
\sum_{J \neq K} 
\left[ 
\mbox{Re} \left( W_{\alpha J}^* W_{\beta J} W_{\alpha K} W^{*}_{\beta K} \right) 
\cos ( \Delta_{K} - \Delta_{J}  ) x 
+ \mbox{Im} \left( W_{\alpha J}^* W_{\beta J} W_{\alpha K} W^{*}_{\beta K} \right)
\sin ( \Delta_{K} - \Delta_{J} ) x 
\right]
\nonumber\\
&+& 
2 \sum_{j=1}^{3} \sum_{K=4}^{3+N}
\left[
\mbox{Re} 
\left( U_{\alpha j}^{*} U_{\beta j} W_{\alpha K} W^{*}_{\beta K} \right) 
\cos ( \Delta_{K} - \Delta_{j} ) x 
+ \mbox{Im} \left( U_{\alpha j}^{*} U_{\beta j} W_{\alpha K} W^{*}_{\beta K} \right) 
\sin ( \Delta_{K} - \Delta_{j} ) x 
\right]. 
\nonumber \\
\label{P-beta-alpha-vac}
\end{eqnarray}
We notice that the last two lines vanish after averaging over energy resolution, as discussed in section~\ref{sec:low-E-unitarity-violation}. Then, we obtain the expressions of oscillation probabilities in our $(3 + N)$ model in vacuum.
In the appearance channel, $\alpha \neq \beta$, it reads 
\begin{eqnarray}
P(\nu_\beta \rightarrow \nu_\alpha) &=& 
\mathcal{C}_{\alpha \beta} + 
\left| \sum_{j=1}^{3} U_{\alpha j} U^{*}_{\beta j} \right|^2 - 
2 \sum_{j \neq k} 
\mbox{Re} 
\left( U_{\alpha j} U_{\beta j}^* U_{\alpha k}^* U_{\beta k} \right) 
\sin^2 \frac{ ( \Delta_{k} - \Delta_{j} ) x  }{ 2 }
\nonumber\\
&-&
\sum_{j \neq k} \mbox{Im} 
\left( U_{\alpha j} U_{\beta j}^* U_{\alpha k}^* U_{\beta k} \right) 
\sin ( \Delta_{k} - \Delta_{j} ) x, 
\label{P-beta-alpha-ave-vac}
\end{eqnarray}
and in the disappearance channel
\begin{eqnarray}
P(\nu_\alpha \rightarrow \nu_\alpha) = 
\mathcal{C}_{\alpha \alpha } + 
\left( \sum_{j}^{3} \vert U_{\alpha j} \vert^2 \right)^2 
- 4 \sum_{ k > j }^{3}  \vert U_{\alpha j} \vert^2 \vert U_{\alpha k} \vert^2 
\sin^2 \frac{ ( \Delta_{k} - \Delta_{j} ) x  }{ 2 }, 
\label{P-alpha-alpha-ave-vac}
\end{eqnarray}
where
\begin{eqnarray} 
\mathcal{C}_{\alpha \beta} \equiv 
\sum_{J=4}^{3+N}
\vert W_{\alpha J} \vert^2 \vert W_{\beta J} \vert^2, 
\hspace{10mm}
\mathcal{C}_{\alpha \alpha } \equiv 
\sum_{J=4}^{3+N} 
\vert W_{\alpha J} \vert^4.
\label{Cab-Caa}
\end{eqnarray}

One should notice that after averaging over high-frequency sterile oscillations, the expressions in (\ref{P-beta-alpha-ave-vac}) and (\ref{P-alpha-alpha-ave-vac}) have terms which look like the ``zero-distance flavor transition''. But, it cannot be the correct interpretation because the averaging procedure (even though it is on energy spectrum) inherently contains certain distance scale to observe destructive interference which leads to cancellation of oscillatory behavior.

The expression of the oscillation probabilities in (\ref{P-beta-alpha-ave-vac}) and (\ref{P-alpha-alpha-ave-vac}) look similar to the ones in the standard three-flavor mixing. But, there are two important differences: 

\begin{itemize}

\item
The active space mixing matrix $U$ is not unitary, 

\item
There is a probability leaking term to the sterile neutrino sector, $\mathcal{C}_{\alpha \beta}$ in (\ref{P-beta-alpha-ave-vac}) and $\mathcal{C}_{\alpha \alpha }$ in (\ref{P-alpha-alpha-ave-vac}). 

\end{itemize}
\noindent
The former is a common feature of the theories in which unitarity is
violated in active neutrino subspace. 
In the unitary case the second term in 
(\ref{P-beta-alpha-ave-vac}) is $\delta_{\alpha \beta}$.
On the other hand, the second point above, the existence of probability leaking term, is the characteristic feature of the low-scale unitarity violation. 
However, the term is omitted in the expression of the oscillation probability in the literature, e.g., in refs.~\cite{Parke:2015goa,Qian:2013ora}, and was considered only for some specific models of sterile 
neutrinos, e.g., in \cite{Maltoni:2007zf,Li:2015oal}.

Does the leaking term introduce a heavy model-dependence into the prediction by our $(3+N)$ model? The answer is no: though it indeed displays some sterile sector model dependence, it is only a mild one. That is, the term can be treated as the channel dependent constant $\mathcal{C}_{\alpha \beta}$ when this formula is used to analyze leptonic unitarity violation in vacuum.

We emphasize that the clearest evidence for low-scale unitarity
violation is the demonstration of existence of probability leaking
constant 
$\mathcal{C}_{\alpha \beta} \equiv \sum_{J=4}^{3+N} \vert W_{\alpha J}\vert^2 \vert W_{\beta J} \vert^2$. 
Unfortunately, it would not be easy to carry out for the two reasons: (1) the term is small in size because it is the fourth order in unitarity-violating elements $W_{\alpha J}$, and (2) it is just a constant term and hence it could be confused by the uncertainty in the flux normalization of neutrino beams.

Apart from the probability leaking term $\mathcal{C}_{\alpha \beta}$ ($\alpha = \beta$, $\alpha \neq \beta$), our formulas agree with those of ref.~\cite{Escrihuela:2015wra}. On the other hand, the oscillation probability formulas in ref.~\cite{Antusch:2006vwa} have extra normalization factor. Therefore, it looks like they do not agree with each other although they are both dealing with high-scale unitarity violation. But, since the  normalization factor cancels against those included in the neutrino cross sections they are consistent, if the probability formulas in \cite{Escrihuela:2015wra} are understood as the ones after the cancellation.

\subsection{$(3+N)$ state space unitarity and constraint on probability leaking term}
\label{sec:constraint}

In our three active plus $N$ sterile neutrino model, unitarity is obeyed in the whole $(3+N)$ state space, ${\bf U} {\bf U}^{\dagger} = {\bf U}^{\dagger} {\bf U} = {\bf 1}$. It takes the form in the active $3 \times 3$ subspace 
\begin{eqnarray}
U U^{\dagger} + W W^{\dagger} = 1_{3 \times 3}, 
\hspace{10mm}
U^{\dagger} U + Z^{\dagger}Z = 1_{3 \times 3}.
\label{eqn:unitarity}
\end{eqnarray}
The first relation in (\ref{eqn:unitarity}) implies 
that size of the probability leaking terms, $\mathcal{C}_{\alpha \beta}$ or $\mathcal{C}_{\alpha \alpha}$, and the size of unitarity violation in active space $U$ matrix are related to each other. 

In fact, it is easy to derive the upper and lower bounds on 
$\mathcal{C}_{\alpha \beta} = \sum_{J=4}^{3+N} \vert W_{\alpha J} \vert^2\vert W_{\beta J} \vert^2$ 
and 
$\mathcal{C}_{\alpha \alpha} = \sum_{J=4}^{3+N} \vert W_{\alpha J} \vert^4$. One can start from 
\begin{eqnarray} 
\left( \sum_{I} \vert W_{\alpha I} \vert^2 \right) 
\left( \sum_{J} \vert W_{\beta J} \vert^2 \right)
&=&
\sum_{J} \vert W_{\alpha J} \vert^2 \vert W_{\beta J} \vert^2 + 
\sum_{I \neq J} \vert W_{\alpha I} \vert^2 \vert W_{\beta J} \vert^2
\hspace{6mm}
(\alpha \neq \beta), 
\nonumber \\
\left( \sum_{J} \vert W_{\alpha J} \vert^2 \right)^2 
&=&
\sum_{J=1}^{N} \vert W_{\alpha J} \vert^4 +
\sum_{I \neq J} \vert W_{\alpha I} \vert^2 \vert W_{\alpha J} \vert^2.  
\end{eqnarray}
Since the last terms are non-negative we obtain the upper bounds\footnote{
As pointed out in ref.~\cite{Parke:2015goa}, for $\alpha \neq \beta$ and $i \neq j$ cases respectively, there are two relevant bounds that can
be obtained by applying Cauchy-Schwartz inequalities to unitarity constraints \eqref{eqn:unitarity}:
$|\sum_{i=1}^3 U_{\alpha i} U_{\beta i}^*|^2 \leq \left( 1- \sum_{j=1}^{3} \vert U_{\alpha j} \vert^2 \right) 
\left( 1- \sum_{j=1}^{3} \vert U_{\beta j} \vert^2 \right)$
and $|\sum_{\alpha=e}^\tau U_{\alpha i} U_{\alpha j}^*|^2 \leq \left( 1- \sum_{\alpha=e}^{\tau} \vert U_{\alpha i} \vert^2 \right) 
\left( 1- \sum_{\alpha=e}^{\tau} \vert U_{\alpha j} \vert^2 \right)$. These bounds are relevant when studying 
neutrino appearance $\nu_\alpha \to \nu_\beta$.
}
\begin{eqnarray}
\mathcal{C}_{\alpha \beta} &\leq& 
\left( 1- \sum_{j=1}^{3} \vert U_{\alpha j} \vert^2 \right) 
\left( 1- \sum_{j=1}^{3} \vert U_{\beta j} \vert^2 \right)
\hspace{6mm}
(\alpha \neq \beta), 
\nonumber \\
\mathcal{C}_{\alpha \alpha} &\leq& 
\left( 1- \sum_{j=1}^{3} \vert U_{\alpha j} \vert^2 \right)^2. 
\label{eqn:H-max-ab}
\end{eqnarray}
The lower bound is slightly nontrivial, but they are derived in appendix~\ref{sec:bound-C}: 
\begin{eqnarray}
\mathcal{C}_{\alpha \beta} &\geq& 
\frac{ 1 }{ N } 
\left( 1- \sum_{j=1}^{3} \vert U_{\alpha j} \vert^2 \right) 
\left( 1- \sum_{j=1}^{3} \vert U_{\beta j} \vert^2 \right)
\hspace{6mm}
(\alpha \neq \beta), 
\nonumber \\
\mathcal{C}_{\alpha \alpha} &\geq& 
\frac{ 1 }{ N } 
\left( 1- \sum_{j=1}^{3} \vert U_{\alpha j} \vert^2 \right)^2. 
\label{eqn:H-min-ab}
\end{eqnarray}
The lower bounds depend on $N$, and therefore they are sterile-sector model dependent. But, since the upper bounds are more restrictive, as we will see in the analysis in section~\ref{sec:JUNO}, we assume the least restrictive case, $N=\infty$ there. 

Using (\ref{eqn:H-max-ab}) and (\ref{eqn:H-min-ab}), and the fact that $(1 - \sum_{i=1}^{3} |U_{\alpha i}|^2)$ and $\mathcal{C}_{\alpha \alpha}$ are both positive, one can derive the bound 
$\sqrt{ \mathcal{C}_{\alpha \alpha} } \leq  (1 - \sum_{i=1}^{3} |U_{\alpha i}|^2) \leq \sqrt{ N \mathcal{C}_{\alpha \alpha} }$. Suppose that the analysis of future experimental data indicates unitarity violation with nonzero value of $\mathcal{C}_{\alpha \alpha}$ and $(1 - \sum_{i=1}^{3} |U_{\alpha i}|^2)$. If the data shows $(1 - \sum_{i=1}^{3} |U_{\alpha i}|^2) = \sqrt{M \mathcal{C}_{\alpha \alpha} }$. Then, the $(3+N)$ space unitary model with $N < M$ is excluded.

\subsection{Summarizing our method of testing leptonic unitarity} 
\label{sec:summary}

Now, we can summarize our method of testing our $(3+N)$ model of low-scale unitarity violation in vacuum: 

\vspace{3mm}
We fit the data by using the two ansatz: (1) the standard three-flavor mixing with unitary mixing matrix $U_\text{PDG}$ \cite{Agashe:2014kda}, and (2) the expressions of the oscillation probabilities in (\ref{P-beta-alpha-ave-vac}) and (\ref{P-alpha-alpha-ave-vac}), with the non-unitary $U$ matrix and the probability leaking terms $\mathcal{C}_{\alpha \beta}$ and/or $\mathcal{C}_{\alpha \alpha }$. In the latter fit, it is important to place the constraints (\ref{eqn:H-max-ab}) on $\mathcal{C}_{\alpha \beta}$ and $\mathcal{C}_{\alpha \alpha }$. In section~\ref{sec:JUNO} we present an analysis of simulated JUNO data within our formalism.

\vspace{3mm}
One can think of various features of the fit results that can be obtained in this way. To discuss possible implications, let us assume for conceptual clarity that a set of super-high precision measurement were done by experiments with perfectly controlled neutrino beam.

\begin{itemize}

\item
If the fit results using (1) the standard three-flavor mixing, and (2) the $(3+N)$ model reveal only small difference between them, it is an indication of absence of unitarity violation. One can obtain quantitative bounds on how severely unitarity violation is constrained. 

\item
If the fit revealed a discrepancy between (1) and (2), it is an indication of unitarity violation. It is likely that the first indication of unitarity violation comes from nonzero values of $1 - \sum_{i=1}^{3} \vert U_{\alpha i} \vert^2$ ($\alpha = e, \mu, \tau$) in the disappearance channels, and/or $\left| \sum_{j=1}^{3} U_{\alpha j} U^{*}_{\beta j} \right|$ in the appearance channels. They are both of the order of $W^2$.

\item
If the measurement is sufficiently accurate to detect nonzero values of $\mathcal{C}_{\alpha \beta}$ ($\alpha \neq \beta$ and/or $\alpha = \beta$) of the order of $W^4$, in addition to nonzero $1 - \sum_{i=1}^{3} \vert U_{\alpha i} \vert^2$ and/or $\left| \sum_{j=1}^{3} U_{\alpha j} U^{*}_{\beta j} \right|$, it is a hint for low-scale unitarity violation. 

\item
If the fit revealed a discrepancy between (1) and (2), indicating unitarity violation, and the fit results of $\mathcal{C}_{\alpha \beta}$ ($\alpha \neq \beta$ and/or $\alpha = \beta$) is outside the region allowed by the constraints (\ref{eqn:H-max-ab}). Nonvanishing $\mathcal{C}_{\alpha \beta}$ suggests unitarity violation at low energies, which however implies that either both the conditions \eqref{eq:decoherence_x} and \eqref{eq:decoherence_E} are not satisfied or the scenario cannot be described by our $(3+N)$ space unitary model. 

\end{itemize}
\noindent
The final consistency check for proving low-scale unitarity violation in the third case above is to verify 
(i) the consistency between the magnitudes of $\mathcal{C}_{\alpha \beta}$ ($\sim W^4$) and $1 - \sum_{i=1}^{3} \vert U_{\alpha i} \vert^2$ and/or $\left| \sum_{j=1}^{3} U_{\alpha j} U^{*}_{\beta j} \right|$ ($\sim W^2$), and 
(ii) over-all consistency between deviation of unitarity of $U$ matrix and the size of $W$ matrix expected from the $(3+N)$ space unitarity (\ref{eqn:unitarity}). 
We note that the relative magnitudes of 
$\mathcal{C}_{\alpha \beta}$ and $1 - \sum_{i=1}^{3} \vert U_{\alpha i} \vert^2$ (or $\left| \sum_{j=1}^{3} U_{\alpha j} U^{*}_{\beta j} \right|$) is also enforced by the upper and lower bounds (\ref{eqn:H-max-ab}) and (\ref{eqn:H-min-ab}), and therefore the property is in the heart of the $(3+N)$ space unitary model.

A clarifying remark is in order: 
In the appearance oscillation probability, (\ref{P-beta-alpha-ave-vac}), $\left| \sum_{j=1}^{3} U_{\alpha j} U^{*}_{\beta j} \right|$ comes in as squared and the term is of the same order $\sim W^4$ as the leaking term $\mathcal{C}_{\alpha \beta}$. Therefore, one might think that the better accuracy may not be expected for $\left| \sum_{j=1}^{3} U_{\alpha j} U^{*}_{\beta j} \right|$. The statement above that ``$\left| \sum_{j=1}^{3} U_{\alpha j} U^{*}_{\beta j} \right|$ is the first indicator of unitarity violation'' really means that the non-unitary $U$ matrix elements are determined mostly by the $x/E$ dependent oscillation terms and it determines (or strongly constrains) $\left| \sum_{j=1}^{3} U_{\alpha j} U^{*}_{\beta j} \right|$, and in this way a better accuracy is expected for $\left| \sum_{j=1}^{3} U_{\alpha j} U^{*}_{\beta j} \right|$. The similar statement for disappearance channel also follows.

\section{Unitarity violation: Case study using JUNO-like setting and the current constraints  }
\label{sec:JUNO}

In this section we carry out the first test of our framework describing low-scale unitarity violation by applying it to data to be obtained by medium-baseline reactor neutrino experiments. For definiteness we assume the JUNO-like setting as defined below.\footnote{
The similar analysis of simulated JUNO data in the context of leptonic
unitarity test was carried out in ref.~\cite{Qian:2013ora}. See also section 3.3 of \cite{JUNO}. 
}
 
We define our analysis method in section \ref{sec:method} and present the results in section~\ref{sec:result}. During the course of describing the results of our analysis, a comparison with the constraints currently available for the $\nu_{e}$ channel will be done. For the $\nu_{\mu}$ and $\nu_{\tau}$ related channels, we will give a brief overview of the current constraints in section~\ref{sec:constraints}, together with miscellaneous remarks on the $\nu_{e}$ channel. 

In our analysis using the JUNO-like setting, we give special attention to the probability leaking term $\mathcal{C}_{\alpha \alpha}$ ($\alpha=e$) 
in eq.~(\ref{P-alpha-alpha-ave-vac}), as discussed in section~\ref{sec:summary}. Of course, estimation of JUNO's capability of constraining (or probing) non-unitary nature of active space $U$ matrix in the $\nu_{e}$ sector is a very interesting point by itself. 
Yet, we must admit that our analysis using a simple-minded $\chi^2$
 cannot be considered as the real quantitative one. We use the
 expression of disappearance probability $P(\nu_\alpha \rightarrow
 \nu_\alpha)$ ($\alpha=e$) in eq.~(\ref{P-alpha-alpha-ave-vac}) for
 reactor neutrino analysis because it is identical to
 $P(\bar{\nu}_\alpha \rightarrow \bar{\nu}_\alpha)$ assuming CPT invariance.

\subsection{Analysis method}
\label{sec:method}

We basically follow the analysis done in \cite{Abrahao:2015rba} with
some modification and simplification. 
In our statistical analysis, we define the $\chi^2$ function 
which consists of two terms as, 
\begin{eqnarray}
\chi^2 \equiv 
\chi^2_\text{stat} 
+\chi^2_\text{sys}.
\label{eq:chi2}
\end{eqnarray}
In the present analysis we do not take into account any data except for JUNO, not even precision measurement of $\sin^2 \theta_{13}$ by Daya Bay and RENO \cite{An:2015rpe,RENO:2015ksa}, which is expected to be improved to $\sim 3$\% level. On this point, we will make a comment in section~\ref{sec:result}.

Following ~\cite{Ge:2012wj,Capozzi:2013psa}, 
the $\chi^2_\text{stat}$ is defined as, 
\begin{eqnarray}
\chi^2_\text{stat} \equiv 
\int_0^{{E^\text{max}_\text{vis}}} 
dE_\text{vis}
\left(
\frac{
\displaystyle
 \frac{dN^\text{obs}}{dE_\text{vis}}
-f_\text{norm}\sum_{i=\text{reac}}
\frac{dN^\text{fit}_i}{dE_\text{vis}} }
{\displaystyle
\sqrt{ 
\frac{dN^\text{obs}}{ dE_\text{vis}} 
} 
}
\right)^2,
\label{eq:chi2_stat}
\end{eqnarray}
where 
$dN^\text{obs} / dE_\text{vis}$ 
denotes the energy distributions of the observed (simulated) signal, 
and $f_\text{norm}$ is the flux normalization parameter for reactor neutrinos,
to be varied freely subject to the pull term in $\chi^2_\text{sys}$ 
(see below) and we integrate up to $E^\text{max}_\text{vis} = 8$ MeV. 
Due to the lack of space, we do not describe here how to compute 
the event number distribution $dN^\text{obs} / dE_\text{vis}$, 
leaving it to appendix \ref{number-of-events}. 

We consider only one kind of systematic error to take into account
the reactor neutrino flux uncertainty, 
\begin{eqnarray}
\chi^2_\text{sys} 
\equiv 
\left(\frac{1-f_\text{norm} }
{ \sigma_{{f}_\text{norm}} } \right)^2 
\label{eq:chi2_sys},
\end{eqnarray}
and use $\sigma_{f_\text{norm}}$ = 3\% as the reference value,  
assuming progress in understanding of the reactor neutrino flux at the JUNO measurement era. 
Yet, given the current status of simulating reactor neutrino flux, we also examine the case of $\sigma_{f_\text{norm}}$ = 6\% for comparison.

There are five relevant free parameters to be fitted in our analysis, namely, 
$|U_{e1}|^2$, $|U_{e2}|^2$, $\sum_{i=1}^3 |U_{ei}|^2$, $\mathcal{C}_{ee}$
as well as the flux normalization parameter, $f_\text{norm}$.
These five parameters are varied freely 
under the conditions, 
\begin{eqnarray}
\sum_{i=1}^3 |U_{ei}|^2 \le 1, \hspace{10mm}
0 \le \mathcal{C}_{ee} \le (1-\sum_{i=1}^3 |U_{ei}|^2)^2,
\label{eq:restrictions}
\end{eqnarray}
as well as with the $\chi^2_\text{sys}$ defined in (\ref{eq:chi2_sys}).
For simplicity, we fix the two mass squared differences as 
$\Delta m^2_{21} = 7.5 \times 10^{-5}$ eV$^2$, 
$\Delta m^2_{31} = 2.46 \times 10^{-3}$ eV$^2$ and consider
only the case of normal mass hierarchy.  
We believe that even if we vary them our results would not change
significantly. 

\begin{figure}[h!]
\begin{center}
\hspace{-18mm}
\includegraphics[bb=0 0 792 600,width=1.1\textwidth]{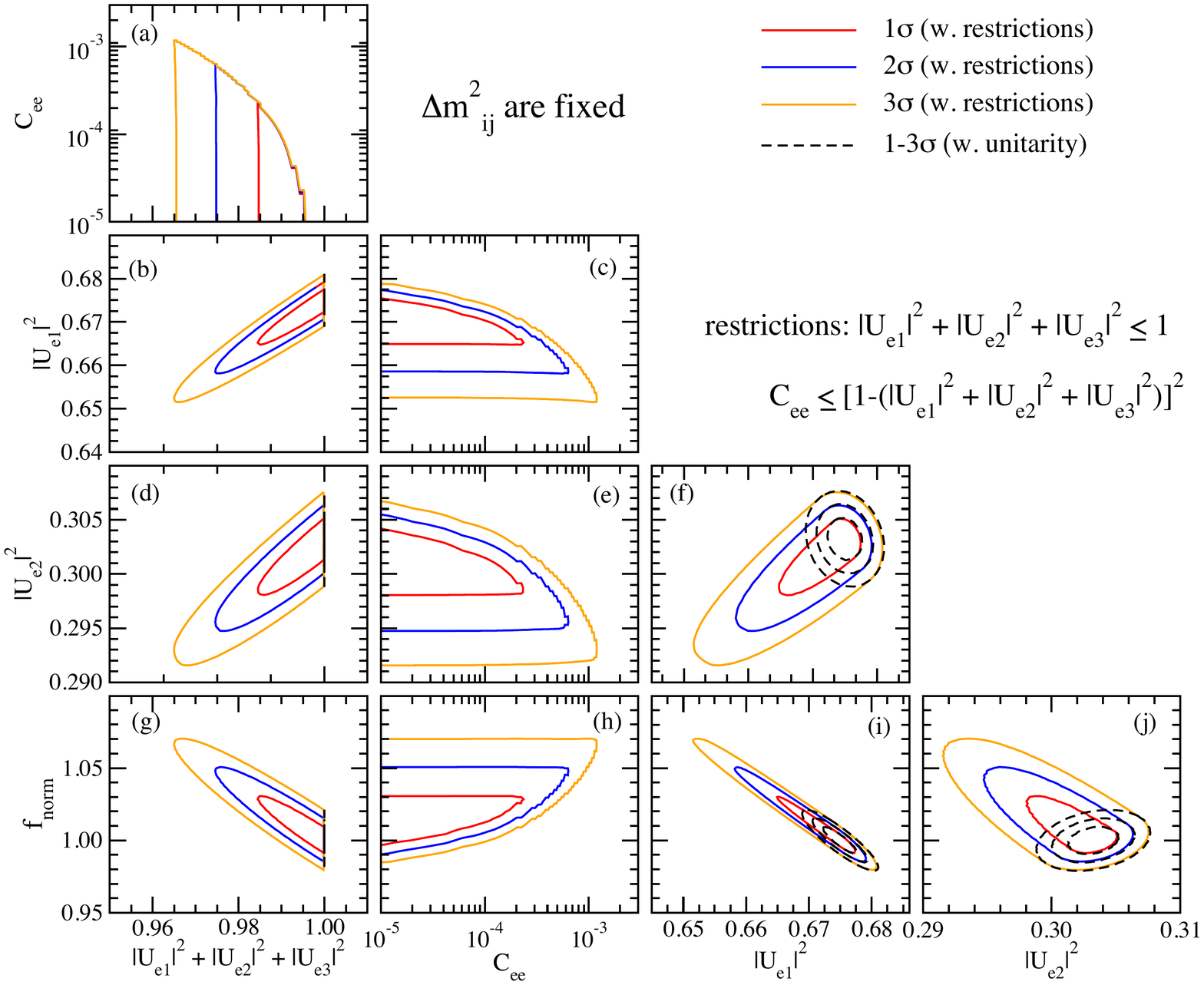}
\end{center}
\vspace{-5mm}
\caption{ 
Regions allowed for the five parameters $|U_{e1}|^2$, $|U_{e2}|^2$, $\sum_{i=1}^3 |U_{ei}|^2$, $\mathcal{C}_{ee}$ and $f_\text{norm}$, are plotted by projecting into each 2 dimensional subspace at 1$\sigma$, 2$\sigma$ and 3$\sigma$ CL. The case of reactor neutrino flux uncertainty of 3\%. 
The colored solid contours are for the cases with unitarity violation under the conditions $\sum_{i=1}^3 |U_{ei}|^2 \le 1$ and $0 \le \mathcal{C}_{ee} \le (1-\sum_{i=1}^3 |U_{ei}|^2)^2$.
The black dashed contours are for the standard unitary case.  }
\label{fig:U-UV-comparison3}
\end{figure}

Using the $\chi^2$ function, we will determine the allowed 
ranges of the five parameters mentioned above, which will 
be projected into two or one dimensional subspace 
by using the conditions, 
\begin{equation}
\Delta \chi^2 \equiv \chi^2 - \chi^2_{{\text{min}}} 
= 2.3,\ 6.18\ \text{and}\ 11.93 \ (1,\ 4\ \text{and} \ 9),
\end{equation}
at 1, 2 and 3 $\sigma$ CL, respectively, for two (one) degrees of freedom. 
The allowed contours obtained by following the above procedure for the
cases of flux normalization uncertainties of 3\% and 6\% are presented
in figures~\ref{fig:U-UV-comparison3} and \ref{fig:U-UV-comparison6}, respectively. 
Since we consider the input which corresponds to the case without
unitarity violation, $\chi^2_\text{min}=0$ by construction as we do not 
take into account the statistical fluctuation in simulating 
the artificial data. 

To understand better the features of the allowed contours in figures~\ref{fig:U-UV-comparison3} and \ref{fig:U-UV-comparison6}, we have also performed the analysis using the same procedure as above but without the constraints (\ref{eq:restrictions}). The results of such analysis with $\sigma_{f_\text{norm}}$ = 3\% are given in figure~\ref{fig:UV-C-NC} in appendix~\ref{sec:C-NC}.

\subsection{Analysis result}
\label{sec:result}

In this section we present the results of our analysis of simulated JUNO data with particular emphasis to the bounds on the parameters, $\mathcal{C}_{ee}$
and $1 - \sum_{i=1}^{3} \vert U_{e i} \vert^2$. A nonzero value of
$\mathcal{C}_{ee}$ implies existence of the low-scale unitarity
violation, distinguishing it from the high-scale unitarity violation. 
Unfortunately, size of $\mathcal{C}_{ee}$ is quite small because it is of the order of $W^4$. While the latter, $1 - \sum_{i=1}^{3} \vert U_{e i} \vert^2$, being of the order of $W^2$, must be the first indicator of unitarity violation. 
We generate the input data without considering unitarity violation 
(corresponding to the standard three flavor scheme) but in the fit, 
we allow non-unitarity, 
in order to determine to what extent a JUNO-like experiment can 
constrain non-unitarity when the data are consistent 
with the standard three flavor scenario. 

\subsubsection{Comparison between the unitary and the non-unitary cases}
\label{sec:U-UV}

\begin{figure}[h!]
\begin{center}
\hspace{-18mm}
\includegraphics[bb=0 0 792 600,width=1.1\textwidth]{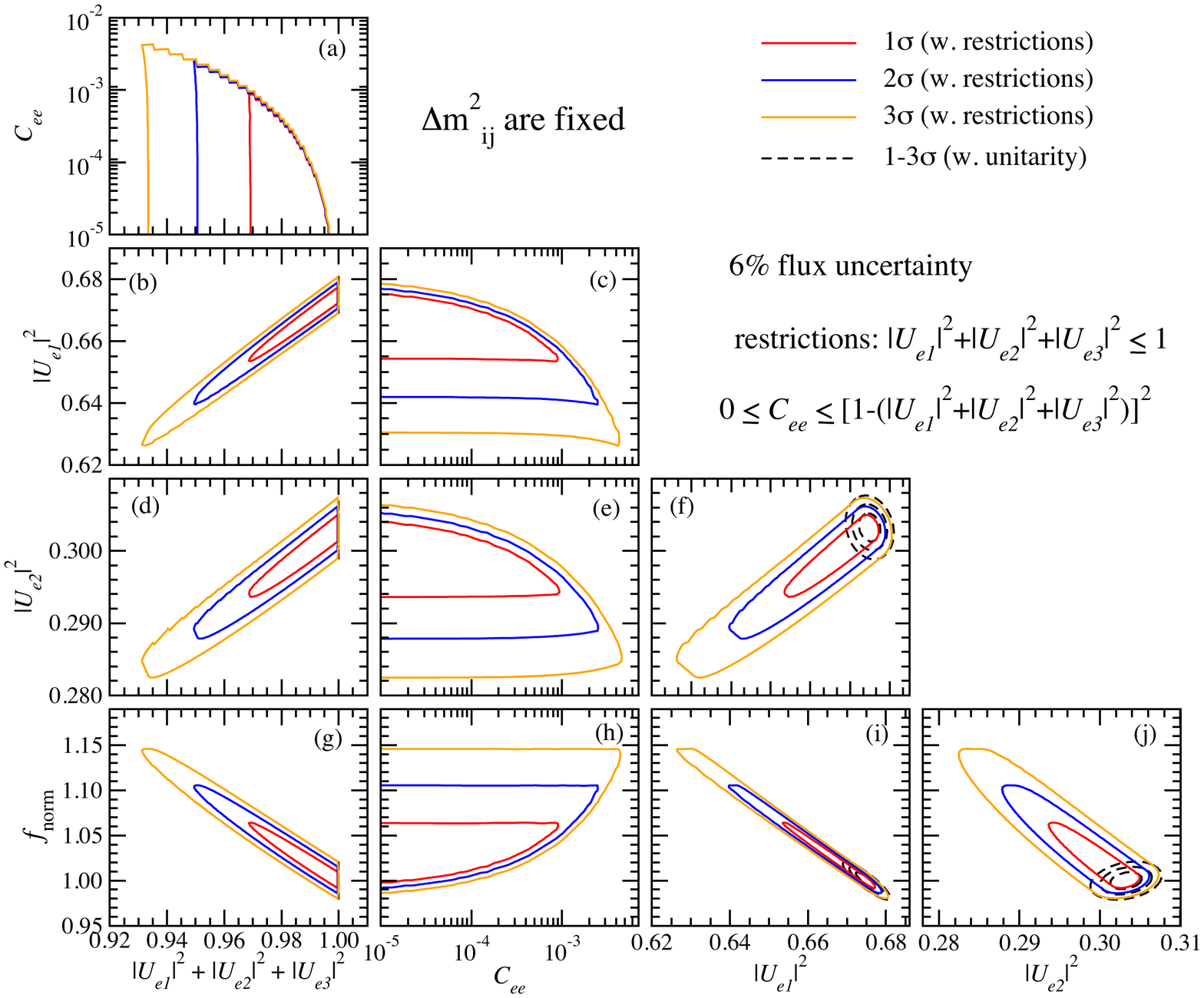}
\end{center}
\vspace{-5mm}
\caption{ 
The same as in figure~\ref{fig:U-UV-comparison3}. 
The case of reactor neutrino flux uncertainty of 6\%. 
\label{fig:U-UV-comparison6}
}
\end{figure}

In figures~\ref{fig:U-UV-comparison3} and \ref{fig:U-UV-comparison6}, 
presented are the allowed regions of $\mathcal{C}_{ee}$, $\sum_{i=1}^{3} \vert U_{ei} \vert^2$, $\vert U_{ei} \vert^2$ ($i=1,2$), and the flux normalization $f_{ \text{norm} }$ projected onto the various two-dimensional spaces at 1, 2, and 3 $\sigma$ CL (each differentiated by colors) obtained with 5 years measurement by JUNO.\footnote{
To be more precise, we consider the total exposure corresponding 
to $5\times 35.8 \times 20 = 3.58\times 10^3$ kt$\cdot$GW$\cdot$yr. 
}
The reactor neutrino flux uncertainty is taken as 3\% and 6\% in 
figures~\ref{fig:U-UV-comparison3} and \ref{fig:U-UV-comparison6},
respectively. 

Alternatively, the allowed regions of unitarity violation parameters $\mathcal{C}_{ee}$, and $1 - (\vert U_{e 1} \vert^2 + \vert U_{e 2} \vert^2 +\vert U_{e 3} \vert^2)$, as well as $f_{ \text{norm} }$, $\vert U_{e1} \vert^2$, and $\vert U_{e2} \vert^2$ at 1 and 3 $\sigma$ CL for 1 degree of freedom are summarized in table~\ref{tab:allowed-range} for the both cases of the reactor flux normalization uncertainties of 3\% and 6\%.

We first concentrate on the former 
(the case for $\sigma_{f_\text{norm}}$=3\%) results given in 
figure \ref{fig:U-UV-comparison3}. 
The colored solid contours are for the cases with unitarity violation, 
while the black dashed contours are for the standard unitary case. 
Since unitarity is preserved in the true (input) simulated data of JUNO, 
the contours obtained with ansatz assuming unitarity violation always 
contain the ones obtained with the standard unitary ansatz.
%
\begin{table}[h!]
\begin{center}
\caption{
Ranges allowed at 1$\sigma$ and 3$\sigma$ CL
of five parameters $|U_{e1}|^2$, $|U_{e2}|^2$, 
$\sum_{i=1}^3 |U_{ei}|^2$, $\mathcal{C}_{ee}$ and $f_\text{norm}$, 
for one degree of freedom 
for 3\% (second and third columns) and 
6\% (fourth and fifth columns) uncertainties of the reactor flux
 normalization.
}
\label{tab:allowed-range} 
\begin{tabular}{c|cc|cc}
\hline 
parameter & 1$\sigma$ range (3\%) & 3$\sigma$ range (3\%) &  1$\sigma$ range (6\%)  &  3$\sigma$ range (6\%) \\
\hline 
$|U_{e1}|^2$ & [0.668, 0.676] &[0.654, 0.680] & [0.661, 0.676]  & [0.632, 0.680] \\
$|U_{e2}|^2$ & [0.299, 0.304] &[0.293, 0.307] & [0.297, 0.304]  & [0.285, 0.307] \\
$\sum_{i=1}^3 |U_{ei}|^2$ & [0.989, 1] & [0.968, 1] & [0.979, 1] & [0.941, 1] \\
$\mathcal{C}_{ee}$ & [0, $10^{-4}$] &[0,$10^{-3}$] & [0, $4\times 10^{-4}$] & [0, $4\times 10^{-3}$] \\
\hline 
$f_{\rm norm}$ & [0.994, 1.02] & [0.983, 1.063] & [0.994, 1.04] & [0.983, 1.13] \\
\hline 
\end{tabular}
\end{center}
\end{table}


Let us understand some key features of figure~\ref{fig:U-UV-comparison3}. The unitarity violation parameter $1 - \sum_{i=1}^{3} \vert U_{e i} \vert^2$ is determined in strong correlation with the flux normalization $f_\text{norm}$. It enters into the constant term in the probability in eq.~\eqref{P-alpha-alpha-ave-vac} with $\alpha=\beta=e$ as 
\begin{eqnarray}
f_\text{norm} (\mathcal{C}_{ee} + \left\{ 
\vert U_{e 1} \vert^2 + \vert U_{e 2} \vert^2 + \vert U_{e 3} \vert^2 
\right\}^2 ) \simeq 
f_\text{norm} \left\{ 
\vert U_{e 1} \vert^2 + \vert U_{e 2} \vert^2 +\vert U_{e 3} \vert^2 
\right\}^2, 
\end{eqnarray}
where the approximate equality above is justified because of the smallness of $\mathcal{C}_{ee}$ as seen in figure~\ref{fig:U-UV-comparison3}. 
Then, it is natural to expect that $1 - \sum_{i=1}^{3} \vert U_{e i}
\vert^2$ would be constrained to the accuracy of $\sim
1-\sqrt{1-\sigma_{f_\text{norm}}}\sim 0.015$. It seems to be consistent
with figure~\ref{fig:U-UV-comparison3}, and  the results given in 
table~\ref{tab:allowed-range},  $1 -
\sum_{i=1}^{3} \vert U_{e i} \vert^2 \leq $ 0.01 (0.03) at 1$\sigma$
(3$\sigma$) CL for one degree of freedom. 


The probability leaking parameter $\mathcal{C}_{ee}$ is constrained to
be small, $\mathcal{C}_{ee} \lsim 2 \times 10^{-4}$ ($10^{-3}$) at
1$\sigma$ (3$\sigma$) CL in figure~\ref{fig:U-UV-comparison3} with two
degrees of freedom, and $\mathcal{C}_{ee} < 10^{-4}$ ($10^{-3}$) at
1$\sigma$ (3$\sigma$) CL in table~\ref{tab:allowed-range} with one degree of
freedom. The stringent constraints obtained for $\mathcal{C}_{ee}$ can
be understood as coming from the upper bound on $\mathcal{C}_{ee}$ in
eq.~(\ref{eq:restrictions}), which is imposed in the analysis. Using the
above bound on the unitarity violation parameter with one degree of
freedom, $\mathcal{C}_{ee} \le (1-\sum_{i=1}^3 |U_{ei}|^2)^2 = 10^{-4}$
($9 \times 10^{-4}$) at 1$\sigma$ (3$\sigma$) CL. They are quite
consistent with the obtained upper bound on $\mathcal{C}_{ee}$ in
table~\ref{tab:allowed-range}. Noticing that $1 - \sum_{i=1}^{3} \vert
U_{e i} \vert^2$ and $\mathcal{C}_{ee}$ are of the order of $W^2$ and
$W^4$, respectively, it means that the $W$ matrix elements are
constrained to be order $\sim 10\%$
by the JUNO measurement.\footnote{
If all the $W$ matrix elements are equal, it means that $\vert W \vert \leq 0.1 / \sqrt{N}$. 
}

Does inclusion of precision data of $\sin^2 \theta_{13}$ to be obtained by future measurement by Daya Bay and RENO of $\sim 3\%$ level significantly improve the sensitivity to unitarity violation? We believe that the answer is no, and here is the reasoning for our belief. The accuracy of measurement of $\sin^2 \theta_{13}$ in JUNO estimated in \cite{Capozzi:2013psa} is $\simeq 7\%$ level, which implies the accuracy $\delta( \sin^2 \theta_{13} ) = 1.5 \times 10^{-3}$. 
It probably means that in our framework the accuracy of measurement of $\vert U_{e 3} \vert^2$ is $\sim 10^{-3}$, which is an order of magnitude smaller than the 1\% level uncertainty of $1 - \sum_{i=1}^{3} \vert U_{e i} \vert^2$. Furthermore, determination of $1 - \sum_{i=1}^{3} \vert U_{e i} \vert^2$ is very weakly correlated with $\vert U_{e 3} \vert^2$. While $\vert U_{e 3} \vert^2$ is measured by detecting small atmospheric ripples on the long-wavelength solar oscillations, $1 - \sum_{i=1}^{3} \vert U_{e i} \vert^2$ is determined in strong correlation with the flux normalization. 

Therefore, it is important to reduce the flux uncertainty in order to increase the sensitivity to unitarity violation, and improvement of the $\vert U_{e 3} \vert^2$ measurement would have much less impact on it.

To examine the effect of worsen reactor flux normalization uncertainty, we have repeated the same calculation with 6\% error, as given in figure~\ref{fig:U-UV-comparison6}. As one can see from the figure, the over-all features of the correlation between the quantities of interests are unchanged. The extent of prolongation of contours due to the worsen flux uncertainty may be estimated once we understand the one for the unitarity violation parameter $1 - (\vert U_{e 1} \vert^2 + \vert U_{e 2} \vert^2 +\vert U_{e 3} \vert^2)$. Following the same logic as above the accuracy of constraining this parameter is expected to be $\sim 1-\sqrt{1-\sigma_{f_\text{norm}}}\sim 0.03$, which is again consistent with figure~\ref{fig:U-UV-comparison6}.

To know to what extent JUNO can tighten the current constraints on the
$\nu_{e}$ row elements, let us compare our results to the ones obtained
in ref.~\cite{Parke:2015goa}. 
We must remark that the authors of ref.~\cite{Parke:2015goa} assumed 5\% uncertainty of reactor neutrino flux. Whereas we use our results obtained by assuming 3\% uncertainty for comparison.
According to the estimate done in this reference (the fourth equation),
the current uncertainties of $\vert U_{e 1} \vert^2$ and $\vert U_{e 2}
\vert^2$ are 11\% and 18\% at 3$\sigma$ CL, respectively. 
On the other hand, the results of our analysis with JUNO-like setting
shows (see table~\ref{tab:allowed-range}) that at 3$\sigma$ CL 
the uncertainties of $\vert U_{e 1} \vert^2$ and $\vert U_{e 2} \vert^2$ 
are, respectively, 1.9\% and 2.3\%.
It implies a great improvement over the current constraints 
by a factor of $\simeq 6$ (8) for 
$\vert U_{e 1} \vert^2$ ($\vert U_{e 2} \vert^2$).
For the 6\% reactor flux normalization uncertainty, the uncertainties of both of 
$\vert U_{e 1} \vert^2$ and $\vert U_{e 2} \vert^2$ are 3.7\%
implying a factor of $\simeq$ 3 (5) improvement for $\vert U_{e 1}\vert^2$ ($\vert U_{e 2}\vert^2$). 

The current constraint on unitarity violating parameter is given by $1 -
(\vert U_{e 1} \vert^2 + \vert U_{e 2} \vert^2 +\vert U_{e 3} \vert^2)
\leq$ 0.074, as one can read off from Fig.~3 of ref.~\cite{Parke:2015goa}.
Whereas in our JUNO analysis, the unitarity violating parameter for 3\% (6\%) flux normalization uncertainty is constrained to be $1 - (\vert U_{e 1} \vert^2 + \vert U_{e 2} \vert^2 +\vert U_{e 3} \vert^2) \leq 0.032$ (0.059), indicating a modest improvement by a factor of $\simeq 2$ (1.2).
The current constraint on 
$1 - (\vert U_{e 1} \vert^2 + \vert U_{e 2} \vert^2 +\vert U_{e 3}
\vert^2)$ suggests that one could obtain the bound on $\mathcal{C}_{ee}$
as $\mathcal{C}_{ee} \le (0.074)^2 \sim 5.5 \times 10^{-3}$, 
if the analysis were done in the similar way as ours. 
We stress that the bound on $\mathcal{C}_{ee}$ obtained by our 
JUNO analysis is stronger by a factor of 5.5.

Notice, however, that under the assumption that the bound on $\vert U_{e
4} \vert^2$ obtained in the framework of $(3+1)$ model translates into
the one on $1 - (\vert U_{e 1} \vert^2 + \vert U_{e 2} \vert^2 +\vert
U_{e 3} \vert^2)$ in the $(3+N)$ model, the kinematical constraint from
beta decay (neutrinoless double beta decay) is severer than the JUNO
bound for massive sterile neutrinos with masses $m^2_{4} \gsim
10^5~{\rm eV}^2$ ($m^2_{4} \gsim 100~{\rm eV}^2$).\footnote{
The bound from neutrinoless double beta decay is valid only if the neutrinos are Majorana particles.
}
See Fig.~4 in ref.~\cite{deGouvea:2015euy}. 

\subsubsection{Understanding correlations between the parameters }
\label{sec:correlation-Ue12}

One observes that, except for the ones which involve $\mathcal{C}_{ee}$, the allowed contours in the non-unitary case are much wider and expanded to the particular direction, indicating the correlations between the
parameters taken in figure~\ref{fig:U-UV-comparison3}. Let us understand
this feature. For this purpose we call readers attention to the bottom 4
panels (g), (h), (i), and (j) in figure~\ref{fig:U-UV-comparison3}. 
In the left-bottom panel (g), we see that $\vert U_{e1} \vert^2 + \vert U_{e2} \vert^2 + \vert U_{e3} \vert^2$ is restricted to be unity in the unitary case, as it should. Whereas, when unitarity violation is allowed, the contours are expanded into a left-up direction. 
The contour cannot expand to the right because $\vert U_{e1} \vert^2 + \vert U_{e2} \vert^2 + \vert U_{e3} \vert^2$ must be equal to or less than unity by $(3+N)$ space unitarity, eq.~(\ref{eqn:unitarity}). They can extend only to left-up direction because the effect of  decrease of $\vert U_{e1} \vert^2 + \vert U_{e2} \vert^2 + \vert U_{e3} \vert^2$ has to be compensated by increase of the flux normalization $f_{ \text{norm} }$. It then explains the similar behavior of the contours in the panels (i), and (j).\footnote{
Later in this section, we offer an alternative but consistent explanation for these features by using a new representation of the $\bar{\nu}_{e}$ survival probability, eq.~(\ref{eqn:Paa-MNPZ}). }

In the panel (f), when unitarity violation is introduced, the allowed contours prolongate to left-down direction, indicating a positive correlation between $\vert U_{e1} \vert^2$ and $\vert U_{e2} \vert^2$. 
If we assume the positive correlation between $\vert U_{e1} \vert^2$ and $\vert U_{e2} \vert^2$, and taking into account that $|U_{e3}|^2 \ll |U_{e1}|^2, |U_{e2}|^2$, we have the positive correlation between $\vert U_{e1} \vert^2$ and $\vert U_{e1} \vert^2 + \vert U_{e2} \vert^2 + \vert U_{e3} \vert^2$ (between $\vert U_{e2} \vert^2$ and $\vert U_{e1} \vert^2 + \vert U_{e2} \vert^2 + \vert U_{e3} \vert^2$), as indicated in the panel (b) ((d)). 
It almost completes the discussion to understand the features of correlations between the quantities plotted in figure~\ref{fig:U-UV-comparison3}. 

Now, what is left is to understand the reason for positive correlation between $\vert U_{e1} \vert^2$ and $\vert U_{e2} \vert^2$, to which we now turn. 
In fact, it is quite a nontrivial feature to understand: If we run the same simulation without the constraint (\ref{eqn:H-max-ab}), we have a negative correlation between $\vert U_{e1} \vert^2$ and $\vert U_{e2} \vert^2$. See figure~\ref{fig:UV-C-NC} in appendix~\ref{sec:C-NC}. 
Here, we focus on the positive correlation between $\vert U_{e1} \vert^2$ and $\vert U_{e2} \vert^2$ seen in figure~\ref{fig:U-UV-comparison3}, and present a model to understand this feature. In appendix~\ref{sec:C-NC}, we will offer the possible explanation of negative correlation between $\vert U_{e1} \vert^2$ and $\vert U_{e2} \vert^2$ in the case without the constraint. 

We have learned from the results of the analysis that 
$1 - \sum_{i=1}^{3} \vert U_{e i} \vert^2$ and $\mathcal{C}_{ee}$ are consistently constrained to be small so that $W^2 \lsim 10^{-2}$. 
It means that the system is nearly unitary. In the unitary case, it is expected that the JUNO setting has sensitivity to the individual $\Delta m^2_{31}$ and $\Delta m^2_{32}$ waves. Let us suppose that this is the case also in the extended parameter space in our $(3+N)$ model. Then, the suitable representation of $P(\bar{\nu}_e \rightarrow \bar{\nu}_e)$ is given by the non-unitary version of the one derived in ref.~\cite{Minakata:2007tn} ($\alpha=e$ below):
\begin{eqnarray}
&& P(\bar{\nu}_\alpha \rightarrow \bar{\nu}_\alpha) = 
\mathcal{C}_{\alpha \alpha} + 
\left\{ 
\vert U_{\alpha 1} \vert^2 + \vert U_{\alpha 2} \vert^2 +\vert U_{\alpha 3} \vert^2 
\right\}^2 - 
4 \vert U_{\alpha 1} \vert^2 \vert U_{\alpha 2} \vert^2 
\sin^2 \frac{ \Delta m^2_{21} x }{ 4E }
\nonumber \\
&-&
2\vert U_{\alpha 3}\vert^2 \left( \vert U_{\alpha 1} \vert^2 +\vert U_{\alpha 2}\vert^2 \right)
\left[ 
1- \sqrt{ 1- 4 XY \sin^2 \frac{ \Delta m^2_{21} x }{ 4E } }
~\cos \left( \frac{\Delta m^2_{\alpha \alpha} x}{2E} \pm \phi^\alpha_\odot \right) 
\right], 
\label{eqn:Paa-MNPZ}
\end{eqnarray}
where 
\begin{eqnarray}
X \equiv \frac{\vert U_{\alpha 1} \vert^2}{ \vert U_{\alpha 1} \vert^2 +\vert U_{\alpha 2}\vert^2 }, 
\hspace{10mm}
Y \equiv \frac{\vert U_{\alpha 2} \vert^2}{ \vert U_{\alpha 1} \vert^2 +\vert U_{\alpha 2}\vert^2 }, 
\end{eqnarray}
and 
\begin{eqnarray}
\Delta m^2_{\alpha \alpha} &\equiv& X \vert \Delta m^2_{31}  \vert + Y \vert \Delta m^2_{32}  \vert, 
\nonumber \\
\phi^{\alpha}_\odot  &= & 
\arctan \left[ 
\left( X - Y \right) \tan \left( \frac{ \Delta m^2_{21} x }{ 4E } \right) \right] - 
\left( X - Y \right) \left( \frac{ \Delta m^2_{21} x }{ 4E } \right).
\end{eqnarray}
$\phi^\alpha_\odot$ is a slowly varying function of $x/E$ which depends only on the solar parameters, see \cite{Minakata:2007tn}. The $\pm$ sign in front of $\phi^\alpha_\odot$ determines the mass ordering. 

Notice that the function inside the square bracket in
(\ref{eqn:Paa-MNPZ}) determines the way how the $\Delta m^2_{31}$ and
$\Delta m^2_{32}$ waves are superposed, and we assume that the JUNO
setting has the sensitivity to it, as was the case of our simple-minded
analysis described in section~\ref{sec:method} used for the unitary case
\cite{Abrahao:2015rba}. Then, variations of the parameters must render
the fast varying function of $x/E$ inside the square bracket be
invariant, at least approximately. 

To compute the number of events, the probability in eq.~(\ref{eqn:Paa-MNPZ}) should be multiplied by the flux normalization factor $f_\text{norm}$, as mentioned in the previous section. Then, we must analyze the effective probability defined as $P(\bar{\nu}_e \rightarrow \bar{\nu}_e)_\text{eff} \equiv f_\text{norm}\times P(\bar{\nu}_e \rightarrow \bar{\nu}_e)$. 
We now look for the transformations which render 
$P(\bar{\nu}_e \rightarrow \bar{\nu}_e)_\text{eff}$ invariant. 
They are 
\begin{eqnarray}
\vert U_{e i} \vert^2 &\rightarrow& \xi \vert U_{e i} \vert^2
\ \ (i=1,2,3), 
\nonumber \\
f_\text{norm} &\rightarrow& \xi^{-2} f_\text{norm},
\nonumber \\
\mathcal{C}_{ee} &\rightarrow& \xi^{2} \mathcal{C}_{ee}, 
\label{eqn:invariant-transf}
\end{eqnarray}
where $\xi$ is an arbitrary parameter. 
Notice that $X$, $Y$, $\Delta m^2_{ee}$, and
$\phi^{\alpha}_\odot$ are manifestly invariant under
(\ref{eqn:invariant-transf}). The invariance of $P(\bar{\nu}_e
\rightarrow \bar{\nu}_e)_\text{eff}$ under (\ref{eqn:invariant-transf})
implies that the allowed contours can be extended to this
``invariance direction''.  Therefore, $\vert U_{e 1} \vert^2$ and
$\vert U_{e 2} \vert^2$ must be positively correlated with each
other, whereas $\vert U_{e i} \vert^2$ ($i=1,2$) and $f_\text{norm}$ is
negatively correlated. 
The former is consistent with the feature shown in panel (f), and the latter in agreement with the one in panel (g), (i), and (j) in figure~\ref{fig:U-UV-comparison3}. Similarly, $\mathcal{C}_{ee}$ must have positive correlation with $\vert U_{e i} \vert^2$ and negative correlation with $f_\text{norm}$, the feature which, however, does not appear to be seen in figure~\ref{fig:U-UV-comparison3}. 
The most important reason for this is $\mathcal{C}_{ee}$ is essentially determined by the conditions given in eq.~(\ref{eq:restrictions}), as mentioned earlier.\footnote{
But, we must note that the validity of the invariance argument is limited. It breaks down at some point because 
(i) first of all, the invariance under the transformations (\ref{eqn:invariant-transf}) 
is broken for $\chi^2$ by the pull term (\ref{eq:chi2_sys}), and 
(ii) the variables we are dealing with live only in restricted ranges, either by $(3+N)$ space unitarity, or as a result of fitting the data. Therefore, the scaling argument has its own inherent limitation.
For the above mentioned features of correlations which involve $\mathcal{C}_{ee}$, these limitations of our invariance argument may not play a key role, because $\mathcal{C}_{ee}$ is restricted to be very small, $\lsim 10^{-3}$ ($3 \sigma$ CL). 
}

\subsection{The current constraints on unitarity violation}  
\label{sec:constraints}

We start by discussing the constraints obtained on unitarity violation in the $\nu_{\mu}$ and $\nu_{\tau}$ channels. We first focus on the relatively low mass sterile states $m^2_{J} \lsim 10~{\rm eV}^2$, and rely on the results obtained by the authors of ref.~\cite{Parke:2015goa}, because their analysis is based on the $(3+N)$ model. We also check the consistency of the results with those in ref.~\cite{Kopp:2013vaa} keeping in mind that most of the analyses in this reference are done using the $(3+1)$ model.

According to ref.~\cite{Parke:2015goa} (see Fig.~3), the unitarity
violating parameter $1 - (\vert U_{\alpha 1} \vert^2 + \vert U_{\alpha
2} \vert^2 +\vert U_{\alpha 3} \vert^2)$ is constrained to be 
$\leq$ 0.064 and $\leq$ 0.44 at 3$\sigma$ CL for $\alpha = \mu$ and $\tau$, respectively. The constraints are obtained by marginalizing over the sterile neutrino masses $\Delta m^2_{J i} \geq 0.01$ eV$^2$.
The constraints on $\vert U_{\mu 4} \vert^2$ and $\vert U_{\tau 4}
\vert^2$ are obtained in ref.~\cite{Kopp:2013vaa} 
(see Fig.~4 of this reference). 
The results can roughly be summarized as $\vert U_{\mu 4} \vert^2 \lsim$
$(1-3) \times 10^{-2}$ for $1~{\rm eV}^2 \lsim \Delta m^2_{41} \lsim 10~{\rm eV}^2$, and $\vert U_{\mu 4} \vert^2 \lsim (3-6) \times 10^{-2}$ for $0.1~{\rm eV}^2 \lsim \Delta m^2_{41} \lsim 1~{\rm eV}^2$. 
The constraints on $\vert U_{\tau 4} \vert^2$ is much milder, when the case of worst phases is taken, $\vert U_{\tau 4} \vert^2 \lsim 0.42$ for the entire region of $\Delta m^2_{41}$ quoted above. The bound on $\vert U_{\alpha 4} \vert^2$ derived in the framework of $(3+1)$ model may be interpreted as the one for $1 - (\vert U_{\alpha 1} \vert^2 + \vert U_{\alpha 2} \vert^2 +\vert U_{\alpha 3} \vert^2)$ in the context of $(3+N)$ model. If we take this interpretation the both results are consistent with each other. We notice that there is an ample room for improvement for the bound on unitarity violation in the $\nu_{\tau}$ channel. 

With regard to the constraints on each individual $\vert U_{\mu i} \vert^2$, the fourth equation of [5]
tells us that 
$0.044 \leq \vert U_{\mu 1} \vert^2 \leq 0.29$ (74\%), 
$0.18 \leq \vert U_{\mu 2} \vert^2 \leq 0.49$ (46\%), and 
$0.37 \leq \vert U_{\mu 3} \vert^2 \leq 0.62$ (25\%) at 3$\sigma$ CL, where the numbers inside parentheses are for percent errors assuming the symmetric errors. 
The similar constraints on $\vert U_{\tau i} \vert^2$ ($i=1,2,3$) are: 
$0.032 \leq \vert U_{\tau 1} \vert^2 \leq 0.34$ (82\%), 
$0.14 \leq \vert U_{\tau 2} \vert^2 \leq 0.52$ (56\%), and 
$0.16 \leq \vert U_{\tau 3} \vert^2 \leq 0.61$ (58\%) at 3$\sigma$ CL. 

If the sterile states are more massive, $m^2_{J} \gsim 10~{\rm eV}^2$, the kinematical constraints in beta and meson decays play more important role.
As we mentioned in section~\ref{sec:U-UV}, the kinematical constraint from neutrinoless double beta decay plays an important role for massive sterile neutrinos, 
$\vert U_{e 4} \vert^2 \lsim 10^{-3}$ for $m^2_{4} \sim 1~\mbox{keV}^2$ to $\vert U_{e 4} \vert^2 \lsim 10^{-6}$ for $m^2_{4} \sim 1~\mbox{MeV}^2$~\cite{deGouvea:2015euy}. 
However, no constraint on $\vert U_{\mu 4} \vert^2$ and $\vert U_{\tau 4} \vert^2$ arizes for the mass range $m_{4} \leq 1~{\rm MeV}$ in which we are interested in the context of low-scale unitarity violation, according to the $(3+1)$ model analysis in \cite{deGouvea:2015euy}.

The neutrino oscillation experiments can constrain the sterile mixing parameters for relatively high mass sterile states, $m^2_{J} \gsim 10~{\rm eV}^2$. Assuming an additional sterile state with $10\,{\rm eV} \lesssim m_4 \lesssim 1\,{\rm MeV}$, the KARMEN experiment constrains $4 |U_{e 4}|^2|U_{\mu 4}|^2 < 1.3 \times 10^{-3}$ at 90 \% CL~\cite{Armbruster:2002mp} while the FNAL-E531 experiment constrains $4 |U_{\mu 4}|^2|U_{\tau 4}|^2 \lsim 4 \times 10^{-3}$ and $4 |U_{e 4}|^2|U_{\tau 4}|^2 \lsim 0.2$ at 90 \% CL~\cite{Ushida:1986zn}. We must note, however, that the precise translation from the constraints obtained by using the $(3+1)$ model to the ones obtainable by our generic (3+$N$) model requires great care. In particular, it is mandatory but is highly nontrivial task for the constraints from accelerator appearance experiments mentioned above.  

For unitarity violation at high scales, due to the SM $SU(2)$ gauge invariance, the constraints coming from the charged lepton sector must also be considered \cite{Antusch:2006vwa}. While we do not describe them here the interested readers are advised to refer to, for example, refs.~\cite{Antusch:2006vwa, Escrihuela:2015wra,Antusch:2014woa,Fernandez-Martinez:2016lgt} and the ones quoted therein. 

\section{Structure of CP violation in the $(3 + N)$ space unitary model}
\label{sec:CP-violation}

As in the preceding section, we can use the formulas for $P(\nu_\mu \rightarrow \nu_e)$ and $P(\nu_\mu \rightarrow \nu_\mu)$ given in (\ref{P-beta-alpha-ave-vac}) and (\ref{P-alpha-alpha-ave-vac}) ($\beta=\mu, \alpha=e$ etc.) to do unitarity test in the accelerator neutrino experiments with muon neutrino beam in near vacuum environment. While we postpone this task to future communications, we want to make remarks on structure of CP violation in the active neutrino sector of our $(3+N)$ unitary model. 
We note that some authors addressed the issue of CP phase in theories with non-unitarity. See e.g., \cite{FernandezMartinez:2007ms,Miranda:2016wdr,Ge:2016xya}. Yet, we believe that our discussion below nicely complements those given before. 

The number of CP violating phases in non-unitary $n \times n$ $U$ matrix can be counted by the similar way as in the CKM matrix: It is $2n^2 - n^2 - (2n-1) = (n-1)^2$, in which we have subtracted number of elements $\vert U_{\alpha i} \vert$ and number of phases that can be absorbed into the neutrino wave functions. Hence, four phases exist in the $U$ matrix in our $(3+N)$ model ($n=3$), and it can be parameterized, for example, as 
\begin{eqnarray}
U &=& 
\left[
\begin{array}{ccc}
\vert U_{e 1} \vert & \vert U_{e 2} \vert & \vert U_{e 3} \vert e^{ i \phi_{1} }  \\
\vert U_{\mu 1} \vert e^{ i \phi_{2} } & \vert U_{\mu 2} \vert & \vert U_{\mu 3} \vert  \\
\vert U_{\tau 1} \vert e^{ i \phi_{3} } & \vert U_{\tau 2} \vert e^{ i \phi_{4} } & \vert U_{\tau 3} \vert  \\
\end{array} 
\right]. 
\label{U-parametrize33}
\end{eqnarray}

Using (\ref{P-beta-alpha-ave-vac}) the CP odd combination of the appearance oscillation probabilities is given by 
\begin{eqnarray} 
\Delta P_{\beta \alpha} \equiv 
P(\nu_\beta \rightarrow \nu_\alpha) -
P(\bar{\nu}_\beta \rightarrow \bar{\nu}_\alpha) 
&=& - 4
\sum_{ j > i } J_{\alpha \beta i j} 
\sin \left(\frac{ \Delta m^2_{ji} x }{2E}\right)
\label{Delta-P}
\end{eqnarray}
where we have defined the generalized Jarlskog invariants \cite{Jarlskog:1985ht}
\begin{eqnarray} 
J_{\alpha \beta i j} \equiv 
\mbox{Im}  \left( U_{\alpha i} U_{\beta i}^* U_{\alpha j}^* U_{\beta j} \right). 
\label{jarlskog}
\end{eqnarray}
They are called ``invariants'' because they are invariant under phase redefinition of neutrino fields. Though $J_{\alpha \beta i j}$ is unique, up to sign, in unitary case, the property no longer holds in our $(3 + N)$ space unitary model. But, some properties remain, e.g., antisymmetry: $J_{\alpha \beta i j} = - J_{\beta \alpha i j}$, $J_{\alpha \beta i j} = - J_{\alpha \beta j i}$. It allows us to show some interesting properties of CP odd combination $\Delta P_{\beta \alpha}$.

Multiplying $U_{\alpha j}^* U_{\beta j}$ to the unitarity relation, the first equation in (\ref{eqn:unitarity}), 
\begin{eqnarray} 
\sum_{i} U_{\alpha i} U_{\beta i}^* = 
\delta_{ \alpha \beta } - 
\sum_{I} W_{\alpha I} W_{\beta I}^*
\label{unitarity1}
\end{eqnarray}
and taking imaginary part we obtain the relation 
\begin{eqnarray} 
\sum_{i} J_{\alpha \beta i j} = - 
\mbox{Im}  \left( U_{\alpha j}^* U_{\beta j} W_{\alpha I} W_{\beta I}^* \right) 
\equiv S_{\alpha \beta j}.
\label{unitarity2}
\end{eqnarray}
Because of antisymmetry of $J_{\alpha \beta i j}$ mentioned above we can write $S_{\alpha \beta j}$ as
\begin{eqnarray} 
S_{\alpha \beta 1} &=& J_{\alpha \beta 21} + J_{\alpha \beta 31}, 
\nonumber \\
S_{\alpha \beta 2} &=& J_{\alpha \beta 12} + J_{\alpha \beta 32}, 
\nonumber \\
S_{\alpha \beta 3} &=& J_{\alpha \beta 13} + J_{\alpha \beta 23}, 
\label{S-J-relation}
\end{eqnarray}
from which the relation $S_{\alpha \beta 1} + S_{\alpha \beta 2} +
S_{\alpha \beta 3} =0$ follows. Then, one can easily show that CP odd 
combination $\Delta P_{\beta \alpha}$ can be written as\footnote{
We have used the identity 
$
\sin \left(\frac{ \Delta m^2_{32} x }{2E}\right) - 
\sin \left(\frac{ \Delta m^2_{31} x }{2E}\right) + 
\sin \left(\frac{ \Delta m^2_{21} x }{2E}\right) = 
4 \sin \left(\frac{ \Delta m^2_{32} x }{4E} \right) 
\sin \left(\frac{ \Delta m^2_{31} x }{4E} \right) 
\sin \left(\frac{ \Delta m^2_{21} x }{4E}\right)$. 
By cyclic permutation one can obtain the other forms with 
the first coefficient $J_{\alpha \beta 23}$ or $J_{\alpha \beta 31}$.  
}
\begin{eqnarray} 
\Delta P_{\beta \alpha} &=& 
- 16 J_{\alpha \beta 12} \sin \left(\frac{ \Delta m^2_{32} x }{4E} \right)
\sin \left(\frac{ \Delta m^2_{31} x }{4E} \right) 
\sin \left(\frac{ \Delta m^2_{21} x }{4E}\right)
\nonumber \\
&+& 
4 S_{\alpha \beta 1} \sin \left(\frac{ \Delta m^2_{31} x }{2E} \right)+ 
4 S_{\alpha \beta 2} \sin \left(\frac{ \Delta m^2_{32} x }{2E}\right). 
\label{Delta-P2}
\end{eqnarray}

The form of CP-odd combination $\Delta P_{\beta \alpha}$ in
(\ref{Delta-P2}) is interesting because CP violation effect is
decomposed into two pieces, one unitary-like $x/E$ dependence (first
line), and the other ``unitarity-violating'' $x/E$ dependence (second
line). Of course, the coefficient of the first term receives unitarity
violating effect through non-unitary $U$ matrix elements in $J_{\alpha
\beta 12}$. 
But, it should be possible to disentangle between these two different
$x/E$ dependences by precision measurement of neutrino spectrum in 
the next generation experiments \cite{Abe:2015zbg,Acciarri:2015uup} provided that the non-unitarity effect is sufficiently large enough. 
Presence of the second term would provide with us a clear evidence for unitarity violation, because $S_{\alpha \beta i}$ involves explicitly the $W$ matrix elements which connect the active to sterile sectors.\footnote{
One must be careful so as not to misinterpret our statement. Through unitarity of ${\bf U}$ matrix (\ref{eqn:unitarity}), the $U$ matrix always carries information of $W$ matrix. Therefore, CP odd term is not the only place where we see the effect of non-unitarity. But, $\Delta P_{\beta \alpha}$ is special because an explicit $W$ dependent piece may be singled out, as we emphasized above.
}

To summarize: 
We have shown in near vacuum environments that the structure of CP odd combination of the appearance oscillation probabilities is illuminating enough to allow us to disentangle unitarity violating piece by studying the $x/E$ dependence of the signal. 

\section{Unitarity violation in matter: Matter perturbation theory}
\label{sec:UV-matter-perturbation}

In this paper we have developed a framework describing unitarity
violation at low energies. It utilizes the three active and $N$ sterile
neutrino state space which is assumed to be complete, i.e., $(3 + N)$
space unitarity. The key issue is whether the model can be formulated in
such a way that its prediction is insensitive to the details of the
sterile sector, for example, the sterile neutrino mass spectrum. In
vacuum we have shown that our $(3 + N)$ model satisfies the requirement
if $m_{J}^2 \gtrsim 0.1$ eV$^2$ for $J \ge 4$.
An immediate question is if this feature survives in matter. In this section, we investigate this problem in a restricted framework of leading-order matter effect perturbation theory. We will answer the question in the positive but under the additional requirement, eq.~(\ref{2nd-condition}). 

We note that our approach which relies on matter perturbation theory is not purely academic. The resultant formulas for the disappearance and appearance probabilities, $P(\nu_\mu \rightarrow \nu_\mu)$ and $P(\nu_\mu \rightarrow \nu_e)$, to first order in matter perturbation theory can be utilized in leptonic unitarity test in T2K and T2HK experiments \cite{Abe:2011ks,Abe:2015zbg}. Notice that keeping higher order terms in $W$ is important because the bound obtainable by the ongoing and the next generation experiments may not be so stringent. Therefore, we do not make any assumptions on the size of $W$ matrix elements in this paper (besides $|W| < 1$). 

\subsection{Matter perturbation theory of three active plus $N$ sterile unitary system}
\label{sec:P-theory}

We formulate the matter perturbation theory of $(3+N)$ space unitary model by assuming that $\vert A \vert \ll |\Delta m^2_{31}|$ where $A \equiv 2\sqrt{2} G_F N_e(x) E$, with $G_F$ being the Fermi constant and $N_e(x)$ electron number density in matter, is the Wolfenstein matter potential \cite{Wolfenstein:1977ue}. In deriving the formulas for the oscillation probabilities, for simplicity, we assume charge-neutrality in matter, and take constant number density approximation for electron, proton and neutron. Inclusion of the spatial dependence can be done assuming adiabaticity, but it will not alter the results in a qualitative way.

To discuss neutrino oscillation in matter in the three active plus $N$ sterile neutrino system the matter potential due to neutral current (NC) as well as charged current (CC) interactions must be taken into account. We therefore take the Hamiltonian in the flavor basis as
\begin{eqnarray} 
H = {\bf U}
\left[
\begin{array}{cccccc}
\Delta_{1} & 0 & 0 & 0 & 0 & 0 \\
0 & \Delta_{2} & 0 & 0 & 0 & 0 \\
0 & 0 & \Delta_{3} & 0 & 0 & 0 \\
0 & 0 & 0 & \Delta_{4} & 0 & 0 \\
0 & 0 & 0 & 0 & \cdot \cdot \cdot & 0 \\
0 & 0 & 0 & 0 & 0 & \Delta_{3+N} \\
\end{array}
\right] 
{\bf U}^{\dagger} 
+
\left[
\begin{array}{cccccc}
\Delta_{A} - \Delta_{B} & 0 & 0 & 0 & 0 & 0 \\
0 & - \Delta_{B} & 0 & 0 & 0 & 0 \\
0 & 0 & - \Delta_{B} & 0 & 0 & 0 \\
0 & 0 & 0 & 0 & 0 & 0 \\
0 & 0 & 0 & 0 & \cdot \cdot \cdot & 0 \\
0 & 0 & 0 & 0 & 0 & 0 \\
\end{array}
\right] 
\label{hamiltonian}
\end{eqnarray}
where 
$\Delta_{i(J)} \equiv \frac{ m^2_{i(J)} }{2E} $ 
as before, as defined in eq.~(\ref{Delta-def}) and, 
\begin{eqnarray} 
\Delta_{A}  \equiv \frac{ A }{2E}, 
\hspace{10mm}
\Delta_{B}  \equiv \frac{ B }{2E}.
\label{Delta-ab-def}
\end{eqnarray}
The matter potentials $A$ and $B$, which are respectively due to CC and NC interactions, take the forms and the values as  
\begin{eqnarray} 
A &=&  
2 \sqrt{2} G_F N_e E \approx 1.52 \times 10^{-4} \left( \frac{Y_e \rho}{\rm g.cm^{-3}} \right) \left( \frac{E}{\rm GeV} \right) {\rm eV}^2, 
\nonumber \\
B &=& \sqrt{2} G_F N_n E = \frac{1}{2} \left( \frac{N_n}{N_e} \right) A,
\label{matt-potential}
\end{eqnarray}
where $N_n$ is the neutron number density in matter. 

\subsection{Perturbation theory in vacuum mass eigenstate basis}

To formulate perturbative treatment it is convenient to work with the
vacuum mass eigenstate basis defined as 
$\tilde{\nu} = ({\bf U}^{\dagger}) \nu$, 
in which the Hamiltonian is related to the flavor basis one as 
$\tilde{H} \equiv {\bf U}^{\dagger} H {\bf U} = \tilde{H}_{0} + \tilde{H}_{1}$, where\footnote{
If we choose a different phase convention e.g. $\tilde H_1 = {\bf U}^\dagger \diag(\Delta_A, 0, 0, \Delta_B, ...,\Delta_B){\bf U}$, the $S$ matrix discussed in the following will change but the physical observable (oscillation probability) remains the same, as it must. This is confirmed by an explicit calculation. 
}
\begin{eqnarray}
\hspace{-0.4cm}
\tilde{H}_{0} =  
\left[
\begin{array}{cccccc}
\Delta_{1} & 0 & 0 & 0 & 0 & 0 \\
0 & \Delta_{2} & 0 & 0 & 0 & 0 \\
0 & 0 & \Delta_{3} & 0 & 0 & 0 \\
0 & 0 & 0 & \Delta_{4} & 0 & 0 \\
0 & 0 & 0 & 0 & \cdot \cdot \cdot & 0 \\
0 & 0 & 0 & 0 & 0 & \Delta_{3+N} \\
\end{array}
\right], 
\hspace{3mm}
\tilde{H}_{1} =
{\bf U}^{\dagger} 
\left[
\begin{array}{cccccc}
\Delta_{A} - \Delta_{B} & 0 & 0 & 0 & 0 & 0 \\
0 & - \Delta_{B} & 0 & 0 & 0 & 0 \\
0 & 0 & - \Delta_{B} & 0 & 0 & 0 \\
0 & 0 & 0 & 0 & 0 & 0 \\
0 & 0 & 0 & 0 & \cdot \cdot \cdot & 0 \\
0 & 0 & 0 & 0 & 0 & 0 \\
\end{array}
\right] {\bf U}. 
\label{H-tilde-3+N}
\end{eqnarray}
The $S$ matrix in the flavor basis $S(x)$ is related to the one in the vacuum mass eigenstate basis $\tilde{S}(x)$ as 
\begin{eqnarray} 
S(x) = {\bf U} \tilde{S} (x) {\bf U}^{\dagger}
\label{Smatrix}
\end{eqnarray}
where 
\begin{eqnarray} 
\tilde{S} (x) = T \text{exp} \left[ -i \int^{x}_{0} dx' \tilde{H} (x')  \right].
\label{tilde-Smatrix}
\end{eqnarray}

We calculate perturbatively the elements of $\tilde{S}$ matrix. Toward the goal, we define $\Omega(x)$ as 
$\Omega(x) = e^{i \tilde{H}_{0} x} \tilde{S} (x)$, which obeys the evolution equation 
\begin{eqnarray} 
i \frac{d}{dx} \Omega(x) = H_{1}(x) \Omega(x) 
\label{omega-evolution}
\end{eqnarray}
where 
\begin{eqnarray} 
H_{1}(x) \equiv e^{i \tilde{H}_{0} x} \tilde{H}_{1} e^{-i \tilde{H}_{0} x}. 
\label{def-H1}
\end{eqnarray}
Then, $\Omega(x)$ can be computed perturbatively as 
\begin{eqnarray} 
\Omega(x) &=& 1 + 
(-i) \int^{x}_{0} dx' H_{1} (x') + 
(-i)^2 \int^{x}_{0} dx' H_{1} (x') \int^{x'}_{0} dx'' H_{1} (x'') + \cdot \cdot \cdot,  
\label{Omega-exp}
\end{eqnarray}
where the ``space-ordered'' form in (\ref{Omega-exp}) is essential 
because of the non-commutativity between $H_{1}(x)$ of different locations. 
Having obtained $\Omega(x)$, $\tilde{ S }$ matrix can be written as 
\begin{eqnarray} 
\tilde{ S }(x) = e^{- i \tilde{H}_{0} x} \Omega(x). 
\label{Smatrix-tilde2}
\end{eqnarray}

We calculate $\tilde{ S }$ matrix to first order in matter perturbation
theory. Since $\tilde{H}_{0}$ is diagonal, $e^{ \pm i \tilde{H}_{0} x}$
takes the simple form $\diag \left( e^{ \pm i \Delta_{1} x }, e^{ \pm i
\Delta_{2} x }, e^{ \pm i \Delta_{3} x },  
e^{ \pm i \Delta_{4} x }, \cdot \cdot \cdot, e^{ \pm i \Delta_{3+N} x } \right)$. 
Using eqs.~(\ref{def-H1}) and (\ref{Omega-exp}) respectively, we first determine $H_1$ and then $\Omega$. 
Using (\ref{Smatrix}), the $S$ matrix elements are given by the $\tilde{ S }$ matrix elements as 
\begin{eqnarray} 
S_{\alpha \beta} &=& 
\sum_{i} U_{\alpha i} U^{*}_{\beta i} \tilde{S}_{ii} 
+ \sum_{i \neq j} U_{\alpha i} U^{*}_{\beta j} \tilde{S}_{ij} 
+ \sum_{I, j} W_{\alpha I} U^{*}_{\beta j} \tilde{S}_{I j} 
+ \sum_{i, J} U_{\alpha i} W^{*}_{\beta J} \tilde{S}_{i J}
\nonumber \\ 
&+& 
\sum_{I} W_{\alpha I} W^{*}_{\beta I} \tilde{S}_{II} 
+ \sum_{I \neq J} W_{\alpha I} W^{*}_{\beta J} \tilde{S}_{IJ},
\label{Smatrix-by-Stilde}
\end{eqnarray}
where the expressions of $\tilde{S}$ matrix elements are given in appendix~\ref{sec:tilde-S}. 
If we decompose $S_{\alpha \beta}$ to zeroth and the first order terms, 
$S_{\alpha \beta} = S_{\alpha \beta}^{(0)} + S_{\alpha \beta}^{(1)}$, we obtain
\begin{eqnarray} 
S_{\alpha \beta}^{(0)} &=& 
\sum_{k} U_{\alpha k} U^{*}_{\beta k} 
e^{ - i \Delta_{k} x} 
+ \sum_{K} W_{\alpha K} W^{*}_{\beta K} 
e^{ - i \Delta_{K} x},
\label{S-elements-3+N-0th}
\end{eqnarray}
which is, of course, identical with (\ref{S-elements-3+N-vac}), and
\begin{eqnarray} 
S_{\alpha \beta}^{(1)} &=& 
\sum_{k} U_{\alpha k} U^{*}_{\beta k} 
e^{ - i \Delta_{k} x} 
\left[ - i (\Delta_{A} x) \vert U_{e k} \vert^2 
+ i (\Delta_{B} x) \sum_{\gamma} \vert U_{\gamma k} \vert^2 
\right] 
\nonumber \\ 
&+& 
\sum_{K} W_{\alpha K} W^{*}_{\beta K} 
e^{ - i \Delta_{K} x} 
\left[
-i ( \Delta_{A} x) \vert U_{e K} \vert^2 
+i ( \Delta_{B} x) \sum_{\gamma} \vert W_{\gamma K} \vert^2  
\right] 
\nonumber \\ 
&+& 
\sum_{k \neq l} U_{\alpha k} U^{*}_{\beta l} 
\left[
\Delta_{A}
U^*_{e k } U_{e l } 
- \Delta_{B} \sum_{\gamma} U^*_{\gamma k} U_{\gamma l} 
\right] 
\frac{ e^{ - i \Delta_{l} x} - e^{ - i \Delta_{k} x} }{ ( \Delta_{l} - \Delta_{k} )} 
\nonumber \\ 
&+& 
\sum_{K, l} W_{\alpha K} U^{*}_{\beta l} 
\left[ 
\Delta_{A} 
W^*_{e K } U_{e l} 
- \Delta_{B} \sum_{\gamma} W^*_{\gamma K} U_{\gamma l} 
\right]
\frac{  e^{ - i \Delta_{l} x} - e^{ - i \Delta_{K} x}  }
{ ( \Delta_{l} - \Delta_{K} )} 
\nonumber \\ 
&+& 
\sum_{k, L} U_{\alpha k} W^{*}_{\beta L} 
\left[
\Delta_{A} 
U^*_{e k} W_{e L} 
- \Delta_{B} \sum_{\gamma} U^*_{\gamma k} W_{\gamma L} 
\right]
\frac{  e^{ - i \Delta_{L} x} - 
e^{ - i \Delta_{k} x}  }{ ( \Delta_{L} - \Delta_{k} )} 
\nonumber \\ 
&+& 
\sum_{K \neq L} W_{\alpha K} W^{*}_{\beta L} 
\left[
\Delta_{A}
W^*_{e K} W_{e L} 
- \Delta_{B} \sum_{\gamma} W^*_{\gamma K} W_{\gamma L} 
\right] 
\frac{ e^{ - i \Delta_{L} x} - 
e^{ - i \Delta_{K} x} }{  ( \Delta_{L} - \Delta_{K} ) }.
\label{S-elements-3+N-1st}
\end{eqnarray}

The oscillation probabilities $P(\nu_\beta \rightarrow \nu_\alpha)$ 
in the appearance ($\beta \neq \alpha$) and disappearance  ($\beta = \alpha$) 
channels can be computed to first order in matter perturbation theory as 
\begin{eqnarray} 
P(\nu_\beta \rightarrow \nu_\alpha) 
&=&
\left| S^{(0)}_{\alpha \beta} + S^{(1)}_{\alpha \beta} \right|^2 = 
\left| S^{(0)}_{\alpha \beta} \right|^2 + 
2 \mbox{Re} \left[ \left( S^{(0)}_{\alpha \beta} \right)^{*} S^{(1)}_{\alpha \beta}  \right].
\label{P-beta-alpha}
\end{eqnarray}
Since the zeroth order term in $P(\nu_\beta \rightarrow \nu_\alpha)$ above 
is already given as the vacuum term, eq.~(\ref{P-beta-alpha-vac}), we only compute the first order matter correction terms. The results of $P(\nu_\alpha \rightarrow \nu_\alpha)^{(1)}$ and $P(\nu_\beta \rightarrow \nu_\alpha)^{(1)}$ are given in appendices~\ref{sec:disapp-P} and \ref{sec:app-P}, respectively.

\subsection{Disappearance channels}
\label{sec:disappearance}

For simplicity, we first discuss the oscillation probability in the
disappearance channel. 
Given the zeroth-order term in eq.~(\ref{P-beta-alpha-vac}), 
we focus on the first-order term here. We
present here $P(\nu_\alpha \rightarrow \nu_\alpha)^{(1)}$ after
averaging over energy resolution and dropping the rapidly oscillating terms due to the large mass squared differences which involve sterile neutrinos\footnote{
The averaging out procedure involves not only (\ref{average-out2}) but also 
\begin{eqnarray}
\left\langle 
(\Delta_{A} x) \sin ( \Delta_{k} - \Delta_{J} ) x 
\right\rangle 
&\approx& 
\left\langle 
(\Delta_{A} x) \sin ( \Delta_{K} - \Delta_{J} ) x
\right\rangle \approx 0, 
\label{average-out3} 
\end{eqnarray}
and cosine as well. It is justified because the rapidly oscillating sine
functions are imposed onto monotonic slowly increasing function of $x$. 
This feature arises due to 
$\frac{\vert \Delta_{A} \vert}{ \Delta_{J} } 
\approx \frac{\vert A \vert}{ \Delta m^2_{J k} } 
\approx \frac{\vert A \vert}{\vert \Delta m^2_{J K} \vert} \ll 1$, see eq.~(\ref{rA-def-value}).
},
\begin{eqnarray} 
&& P(\nu_\alpha \rightarrow \nu_\alpha)^{(1)} = 
2 \mbox{Re} \left[ \left( S^{(0)}_{\alpha \alpha} \right)^{*} S^{(1)}_{\alpha \alpha}  \right] 
\nonumber\\ 
&=& 
- 2 \sum_{j \neq k} 
\vert U_{\alpha j} \vert^2 
\vert U_{\alpha k} \vert^2 
\sin ( \Delta_{k} - \Delta_{j} ) x 
\left[ (\Delta_{A} x) \vert U_{e k} \vert^2 - (\Delta_{B} x)  \sum_{\gamma} \vert U_{\gamma k} \vert^2 
\right] 
\nonumber\\ 
&+& 2 
\sum_{j} \sum_{k \neq l} 
\vert U_{\alpha j} \vert^2 
\mbox{Re} \left[
\Delta_{A}
U_{\alpha k} U^{*}_{\alpha l} 
U^*_{e k } U_{e l } 
- \Delta_{B}
U_{\alpha k} U^{*}_{\alpha l} 
\sum_{\gamma} U^*_{\gamma k} U_{\gamma l} 
\right] 
\frac{ \cos ( \Delta_{l} - \Delta_{j} ) x - \cos ( \Delta_{k} - \Delta_{j} ) x }{ ( \Delta_{l} - \Delta_{k} )} 
\nonumber\\ 
&+& 2 
\sum_{j} \sum_{l} \sum_{K} 
\vert U_{\alpha j} \vert^2 
\mbox{Re} \left[ 
\Delta_{A}
W_{\alpha K} U^{*}_{\alpha l} 
W^*_{e K } U_{e l} 
- \Delta_{B}
W_{\alpha K} U^{*}_{\alpha l} 
\sum_{\gamma} W^*_{\gamma K} U_{\gamma l} 
\right]
\frac{ \cos ( \Delta_{l} -  \Delta_{j} ) x 
}{ ( \Delta_{l} - \Delta_{K} )} 
\nonumber\\ 
&-& 2 
\sum_{j} \sum_{k} \sum_{L} 
\vert U_{\alpha j} \vert^2 
\mbox{Re} \left[
\Delta_{A}
U_{\alpha k} W^{*}_{\alpha L} 
U^*_{e k} W_{e L} 
- \Delta_{B}
U_{\alpha k} W^{*}_{\alpha L} 
\sum_{\gamma} U^*_{\gamma k} W_{\gamma L} 
\right] 
\frac{  \cos ( \Delta_{k} - \Delta_{j} ) x }{ ( \Delta_{L} - \Delta_{k} )}, 
 \label{disapp-P-1st-av}
\end{eqnarray}
leaving the full expression before averaging to appendix~\ref{sec:disapp-P}.

We find that the last two terms in (\ref{disapp-P-1st-av}) violate our requirement that the oscillation probability in our $(3+N)$ model to be insensitive to the spectrum of sterile states unless they are smaller than $\mathcal{C}_{ab} \sim {\cal O}(W^4)$ which implies 
\begin{eqnarray} 
\frac{\vert \Delta_{A} \vert}{ ( \Delta_{J}- \Delta_{k} )} = 
\frac{\vert A \vert}{ \Delta m^2_{J k} } 
\ll \vert W \vert^2. 
\label{2nd-condition}
\end{eqnarray}
A severer restriction is not required because these terms are already suppressed by $W^2$ apart from the energy denominator. 
From 
\begin{eqnarray} 
\frac{\vert A \vert}{ \Delta m^2_{J k} } = 2.13 \times 10^{-3} 
\left(\frac{ \Delta m^2_{J k} }{ 0.1 \mbox{eV}^2}\right)^{-1}
\left(\frac{\rho}{2.8 \,\text{g/cm}^3}\right) \left(\frac{E}{1~\mbox{GeV}}\right),
\label{rA-def-value}
\end{eqnarray}
we notice that, unless $W^2$ is extremely small, $W^2 \lsim 10^{-2}$,
the last two terms in (\ref{disapp-P-1st-av}) can be ignored under the
same condition as in vacuum, $\Delta m^2_{J k} \gsim 0.1$ eV$^2$. If we
discuss the region of $W^2$ which is much smaller, 
we need to restrict ourselves to the case of higher mass sterile
neutrinos. If we treat the regime $W^2 \sim 10^{-3}$ ($W^2 \lsim
10^{-n}$), we need to limit to 
$\Delta m^2_{J k} \simeq m^2_{J} \gsim 1$ eV$^2$ ($10^{(n-3)}$ eV$^2$) 
to keep our $(3+N)$ space unitary model insensitive to details of the sterile sector. 

Assuming the further restriction to the sterile mass spectrum such that 
condition \eqref{2nd-condition} is fulfilled, we obtain the final form of the 
first-order matter correction to $P(\nu_\alpha \rightarrow \nu_\alpha)$ as 
\begin{eqnarray} 
&& P(\nu_\alpha \rightarrow \nu_\alpha)^{(1)} = 
- 2 \sum_{j \neq k} 
\vert U_{\alpha j} \vert^2 
\vert U_{\alpha k} \vert^2 
\sin ( \Delta_{k} - \Delta_{j} ) x 
\left[ (\Delta_{A} x) \vert U_{e k} \vert^2 
- (\Delta_{B} x) \sum_{\gamma} \vert U_{\gamma k} \vert^2
\right] 
\nonumber\\ 
&+& 4 
\sum_{j} \sum_{k \neq l} 
\vert U_{\alpha j} \vert^2 
\mbox{Re} \left[
\Delta_{A}
U_{\alpha k} U^{*}_{\alpha l} 
U^*_{e k } U_{e l } 
- \Delta_{B}
U_{\alpha k} U^{*}_{\alpha l} 
\sum_{\gamma} U^*_{\gamma k} U_{\gamma l} 
\right] \nonumber \\
& \times & 
\frac{ \sin^2 \frac{ ( \Delta_{k} - \Delta_{j} ) x }{2} - \sin^2 \frac{ ( \Delta_{l} - \Delta_{j} ) x }{2} }{ ( \Delta_{l} - \Delta_{k} )}.
\label{disapp-P-1st-1-6-strong-av}
\end{eqnarray}
This expression is written in terms of only active space $U$ matrix elements. Therefore, with additional condition on the sterile neutrino mass spectrum given in (\ref{2nd-condition}), the effect of unitarity violation is only through the non-unitarity $U$ matrix to first order in matter perturbation theory. Thus, we find that the most important modification in the oscillation probability due to non-unitarity is in the vacuum expression in the disappearance channel.

\subsection{Appearance channels}
\label{sec:appearance}

Despite the expression of $P(\nu_\beta \rightarrow \nu_\alpha)^{(1)}$ given in appendix~\ref{sec:app-P} is a little cumbersome, it has a simple form after averaging over neutrino energy within the energy resolution and using the condition (\ref{2nd-condition}): 
\begin{eqnarray} 
&& P(\nu_\beta \rightarrow \nu_\alpha)^{(1)} 
\nonumber\\ 
&=& 
2 \sum_{j \neq k} 
\left[
- \mbox{Re} \left( U^*_{\alpha j} U_{\beta j} U_{\alpha k} U^*_{\beta k} \right)  \sin ( \Delta_{k} - \Delta_{j} ) x + \mbox{Im} \left( U^*_{\alpha j} U_{\beta j} U_{\alpha k} U^*_{\beta k} \right) \cos ( \Delta_{k} - \Delta_{j} ) x 
\right]
\nonumber\\ 
&\times&
\left[ (\Delta_{A} x) \vert U_{e k} \vert^2 - 
(\Delta_{B} x) \sum_{\gamma} \vert U_{\gamma k} \vert^2 \right] 
\nonumber\\ 
&+& 
2 \sum_{j} \sum_{k \neq l} 
\mbox{Re} \left[
\Delta_{A} U^{*}_{\alpha j} U_{\beta j} U_{\alpha k} U^{*}_{\beta l} U^*_{e k } U_{e l } 
- \Delta_{B} U^{*}_{\alpha j} U_{\beta j} U_{\alpha k} U^{*}_{\beta l} 
\sum_{\gamma} U^*_{\gamma k} U_{\gamma l} 
\right] 
\nonumber\\ 
&\times& 
\frac{ \cos ( \Delta_{l} - \Delta_{j} ) x - \cos ( \Delta_{k} - \Delta_{j}) x }{ ( \Delta_{l} - \Delta_{k} )} 
\nonumber\\  
&+& 
2 \sum_{j} \sum_{k \neq l} 
\mbox{Im} \left[
\Delta_{A} U^{*}_{\alpha j} U_{\beta j} U_{\alpha k} U^{*}_{\beta l} U^*_{e k } U_{e l } 
- \Delta_{B} U^{*}_{\alpha j} U_{\beta j} U_{\alpha k} U^{*}_{\beta l} 
\sum_{\gamma} U^*_{\gamma k} U_{\gamma l} 
\right] 
\nonumber\\ 
&\times& 
\frac{ \sin ( \Delta_{l} - \Delta_{j} ) x - \sin ( \Delta_{k} - \Delta_{j}) x }{ ( \Delta_{l} - \Delta_{k} )}. 
\label{app-Pba-1st}
\end{eqnarray}
Again, the survived matter correction terms are written in terms of only active space $U$ matrix elements, leaving the important effect of unitarity violation only in the vacuum term. 

The obvious question would be:  
Do the features obtained in the leading order in matter perturbation theory, in particular, the restriction to the sterile masses (\ref{2nd-condition}), prevails to higher orders? A tantalizing feature of 
the sterile mass condition (\ref{2nd-condition}) is that its fulfillment relies on smallness of $A/\Delta m^2_{J k}$ in our present discussion. Therefore, better treatment of the matter effect is necessary to know whether our $(3+N)$ model can be insensitive to details of the sterile sector under reasonably strong matter effect. We hope to return to these questions in the near future.

\section{Conclusions}

In this paper, we have discussed the relationship between low-scale unitarity violation, the one due to new physics at much lower energies than the electroweak scale, and the conventional high-scale unitarity violation. They include (1) presence (absence) of lepton flavor universality in low-scale (high-scale) unitarity violation, and (2) absence (presence) of zero-distance flavor transition in low-scale (high-scale) unitarity violation. In the case of low-scale unitarity violation, it is likely that extension of low energy lepton sector may enrich the features of neutrino mixing and the effects could be detectable by the precision neutrino oscillation experiments. 

To provide a framework for leptonic unitarity test, by embodying such
features of low-scale unitarity violation, we have constructed a
three-active plus $N$-sterile neutrino model which is assumed to be
unitary in the whole ($3+N$) dimensional state space. Presence of the
sterile sector results in non-unitarity in active three neutrino
subspace. Though inside this specific model, we sought the possibility
that the framework is nearly model-independent to better serve unitarity
test. Namely, we require the prediction of the $(3+N)$ model be
insensitive to the properties of the sterile sector, such as the number
of states $N$ and detailed features of the mass spectrum. We have shown
that restriction to the sterile neutrino masses to 
$m^2_{J} \geq 0.1$ eV$^2$ ($J\ge 4$), due to decoherence, is sufficient to achieve the desired properties, under a mild assumption of no accidental degeneracy in the mass spectrum, i.e., 
$|\Delta m^2_{Ja}| \gg |\Delta m^2_{31}|$, or $\gg \Delta m^2_{21}$ 
where $J=4,.., 3+N, a = 1,...,3+N$. 
The characteristic features of unitarity violation, as modeled by our $(3+N)$ space unitary model, are as follows:
\begin{itemize}

\item
the neutrino oscillation probability contains the constant term $\mathcal{C}_{\alpha \beta}$ in (\ref{Cab-Caa}) ($\alpha \neq \beta$ for appearance channels, and $\alpha = \beta$ for disappearance channels), describing the probability leaking into the sterile subspace. 

\item
the mixing matrix in $3 \times 3$ active neutrino subspace is non-unitary. 

\end{itemize}
\noindent 
While the second feature is common to high- and low-scale unitarity violation, the first feature is unique to low-scale unitarity violation. Since probability leaking occurs due to presence of sterile sector which has energies comparable to active neutrinos we suspect that the first feature above is generic in low-scale unitarity violation even outside of our $(3+N)$ model. 

In our $(3+N)$ space unitary model, the first observable which signals non-unitarity would be nonzero values of $1 - \sum_{i=1}^{3} \vert U_{\alpha i} \vert^2$ ($\alpha = e, \mu, \tau$) in the disappearance channels, and/or $\left| \sum_{j=1}^{3} U_{\alpha j} U^{*}_{\beta j} \right|$ in the appearance channels. They are both of the order of $W^2$, where $W$ is the mixing matrix which connects the active and sterile neutrino subspaces. On the other hand, the probability leaking term $\mathcal{C}_{\alpha \beta}$ (see (\ref{Cab-Caa})) is of the order of $W^4$. 
To verify low-scale unitarity violation, finding a nonzero values of $\mathcal{C}_{\alpha \beta}$ would be enough. But, to prove that unitarity violation occurs in the manner predicted by the $(3+N)$ space unitary model, the consistency between order of magnitudes of $\left| \sum_{j=1}^{3} U_{\alpha j} U^{*}_{\beta j} \right|^2$ and $\mathcal{C}_{\alpha \beta}$ ($\alpha \neq \beta$) (and the corresponding quantities in the disappearance channels) must be checked. 

Thus, we have presented a framework for analysis of unitarity violation
in the lepton sector which is suitable for low-scale unitarity
violation. To examine how it works we have analyzed a simulated data of
medium baseline reactor neutrino experiments prepared by assuming a
JUNO-like setting. By analyzing the data with our simple-minded statistical procedure, 
we have shown that the expected superb performance of JUNO would allow us to constrain unitarity
violation and the probability 
leaking parameters as $1 - \sum_{i=1}^{3} \vert U_{e i} \vert^2 \leq 0.01 (0.03)$ and 
$\mathcal{C}_{ee} \lsim 10^{-4}$ ($10^{-3}$) at 1$\sigma$ (3$\sigma$) CL (one degree of freedom), 
respectively, by its 5 years measurement. 

We have also discussed in a qualitative way how to detect unitarity violation in accelerator appearance measurement. Using the antisymmetry property of the generalized Jarlskog invariants we have shown in section~\ref{sec:CP-violation} that the CP odd combination $P(\nu_\beta \rightarrow \nu_\alpha) - P(\bar{\nu}_\beta \rightarrow \bar{\nu}_\alpha)$ can be decomposed into the two terms with different $x/E$ dependences. See eq.~(\ref{Delta-P2}). If measurement of neutrino energy spectra is sufficiently accurate it would be possible to single out the explicit $W$ matrix dependent piece, providing with us a clear evidence for unitarity violation. 

Finally, we have addressed the question of how inclusion of the matter effect alters the nearly model-independent feature of our $(3+N)$ space unitary model. We have learned that if we discuss the region $W^2 \gsim 10^{-2}$ the 
condition on the sterile neutrino masses 
$m^2_{J} \gsim 0.1$ eV$^2$ needed in vacuum is sufficient, 
but if we want to treat case of even smaller $W^2$, $W^2 \lsim 10^{-n}$, restriction to sterile masses to $m^2_{J} \gsim 10^{(n-3)}$ eV$^2$ is necessary for our $(3+N)$ space unitary model be insensitive to details of the sterile sector. Though our treatment in section~\ref{sec:UV-matter-perturbation} is restricted to first order in matter perturbation theory it is perfectly applicable to the analysis for a class of the LBL experiments, for example, T2HK. Clearly the similar discussion must be attempted under environment of larger matter effect that is expected in some of the next generation LBL experiments such as DUNE. 

\acknowledgments

One of the authors (H.M.) thanks Renata Zukanovich Funchal for intriguing conversations which for him prepared the cradle of this work. He thanks Instituto de F\'{\i}sica, Universidade de S\~ao Paulo for the great opportunity of stay under support by Funda\c{c}\~ao de Amparo \`a Pesquisa do Estado de S\~ao Paulo (FAPESP) with grant number 2015/05208-4. C.S.F. is supported by FAPESP under grants 2013/01792-8 and 2012/10995-7.
H.N. was supported by Funda\c{c}\~ao de Amparo \`a Pesquisa do Estado do Rio de Janeiro (FAPERJ) and Conselho Nacional de Ci\^encia e Tecnologia (CNPq).
H.M. and H.N. thank Masashi Yokoyama and the members of his Group in University of Tokyo for their warm hospitality, where part of this work was carried out. 

\newpage 

\appendix

\section{Bounds on the probability leaking term by $(3+N)$ space unitarity}
\label{sec:bound-C}

Here we would like to derive upper and lower bounds on 
$\mathcal{C}_{\alpha\beta} \equiv \sum_{J=4}^{3+N} |W_{\alpha J}|^2|W_{\beta J}|^2$ 
taking into account the constraint from $(3+N)$ space unitarity. First we have the following identity
\begin{eqnarray}
\mathcal{C}_{\alpha\beta} &=& \left( \sum_{I=4}^{3+N} \vert W_{\alpha I} \vert^2\right)
\left( \sum_{J=4}^{3+N} \vert W_{\beta J} \vert^2\right)
- \sum_{I \neq J} \vert W_{\alpha I} W_{\beta J} \vert^2 \nonumber \\ 
&=& \left(1 - \sum_{i=1}^{3} \vert U_{\alpha i} \vert^2\right)
\left(1 - \sum_{i=1}^{3} \vert U_{\beta i} \vert^2\right)
- \sum_{I \neq J} \vert W_{\alpha I} W_{\beta J} \vert^2,
\label{eq:C_identity}
\end{eqnarray}
where in the second line, we have used the unitarity constraint (second relation of \eqref{eqn:unitarity}).
Holding the first term fixed, we can maximize (minimize) $C_{\alpha\beta}$ by minimizing (maximizing) 
the \emph{non-negative} second term. Geometrically, the lengths of vectors $W_\alpha \equiv \{W_{\alpha 1},...,W_{\alpha N}\}$ and $W_\beta \equiv \{W_{\beta 1},...,W_{\beta N}\}$ in $N$-vector space are fixed and  
we are rotating them to find configurations which minimize or maximize $\mathcal{C}_{\alpha\beta}$.
The second term is non-negative and its minimum is zero.\footnote{
There is a unique configuration: $W_{\alpha I}, W_{\beta I} \neq 0$ for one and only one $I$ while the rest are zero. 
}
Hence $\mathcal{C}_{\alpha\beta}$ is bounded from above by
\begin{eqnarray}
\mathcal{C}_{\alpha\beta}^{\rm max} & = & \left(1 - \sum_{i=1}^{3} \vert U_{\alpha i} \vert^2\right)
\left(1 - \sum_{i=1}^{3} \vert U_{\beta i} \vert^2\right).
\label{eq:Cab-max}
\end{eqnarray}
The maximum of the second term in eq.~\eqref{eq:C_identity} occurs when all the elements of $W_\alpha$ and $W_\beta$ are respectively equal, $W_{\alpha I} \equiv v$ and $ W_{\beta J} \equiv w$:\footnote{
If the readers are not convinced by this argument they can derive the same lower bound (\ref{eq:Cab-min}) by using the Lagrange multiplier method in which one considers
\begin{eqnarray}
H \equiv \sum_{J=4}^{3+N} \vert W_{\alpha J} \vert^2 \vert W_{\beta J} \vert^2 
+ \eta \left( 1- \sum_{j=1}^{3} \vert U_{\alpha j} \vert^2 - 
\sum_{J=4}^{3+N} \vert W_{\alpha J} \vert^2 \right) 
+ \xi \left( 1- \sum_{j=1}^{3} 
\vert U_{\beta j} \vert^2 - 
\sum_{J=4}^{3+N} 
\vert W_{\beta J} \vert^2 \right) 
\nonumber 
\label{eqn:H-def}
\end{eqnarray}
and minimize $H$ in terms of $\vert W_{\alpha J} \vert$, $\eta$, and $\xi$.
 }
\begin{eqnarray}
\mathcal{C}_{\alpha\beta}^{\rm min} &=& \left(1 - \sum_{i=1}^{3} \vert U_{\alpha i} \vert^2\right)
\left(1 - \sum_{i=1}^{3} \vert U_{\beta i} \vert^2\right) - N(N-1) v^2 w^2 \nonumber \\
&=& \left(1 - \sum_{i=1}^{3} \vert U_{\alpha i} \vert^2\right)
\left(1 - \sum_{i=1}^{3} \vert U_{\beta i} \vert^2\right) - (N-1)\mathcal{C}_{\alpha\beta}^{\rm min}.
\end{eqnarray}
The second step above follows from the definition $\mathcal{C}_{\alpha\beta}^{\rm min} = N v^2 w^2$. 
Solving for $\mathcal{C}_{\alpha\beta}^{\rm min}$, we have
\begin{eqnarray}
\mathcal{C}_{\alpha\beta}^{\rm min} &=& \frac{1}{N}\left(1 - \sum_{i=1}^{3} \vert U_{\alpha i} \vert^2\right)
\left(1 - \sum_{i=1}^{3} \vert U_{\beta i} \vert^2\right).
\label{eq:Cab-min}
\end{eqnarray}
The bound on $\mathcal{C}_{\alpha \alpha}$ follows by the similar treatment.

\section{Comparison between the non-unitary constrained and constraint-free cases}
\label{sec:C-NC}

We recognized that for better understanding of the correlations between $\vert U_{e1} \vert^2$ and $\vert U_{e2} \vert^2$, and other features of the contours allowed by JUNO data, 
it is worthwhile to examine the case with and without the constraint (\ref{eqn:H-max-ab}) and compare the results of both cases.

\begin{figure}[h!]
\begin{center}
\vspace{-0.3cm}
\hspace{-16mm}
\includegraphics[bb=0 0 792 600,width=1.0\textwidth]{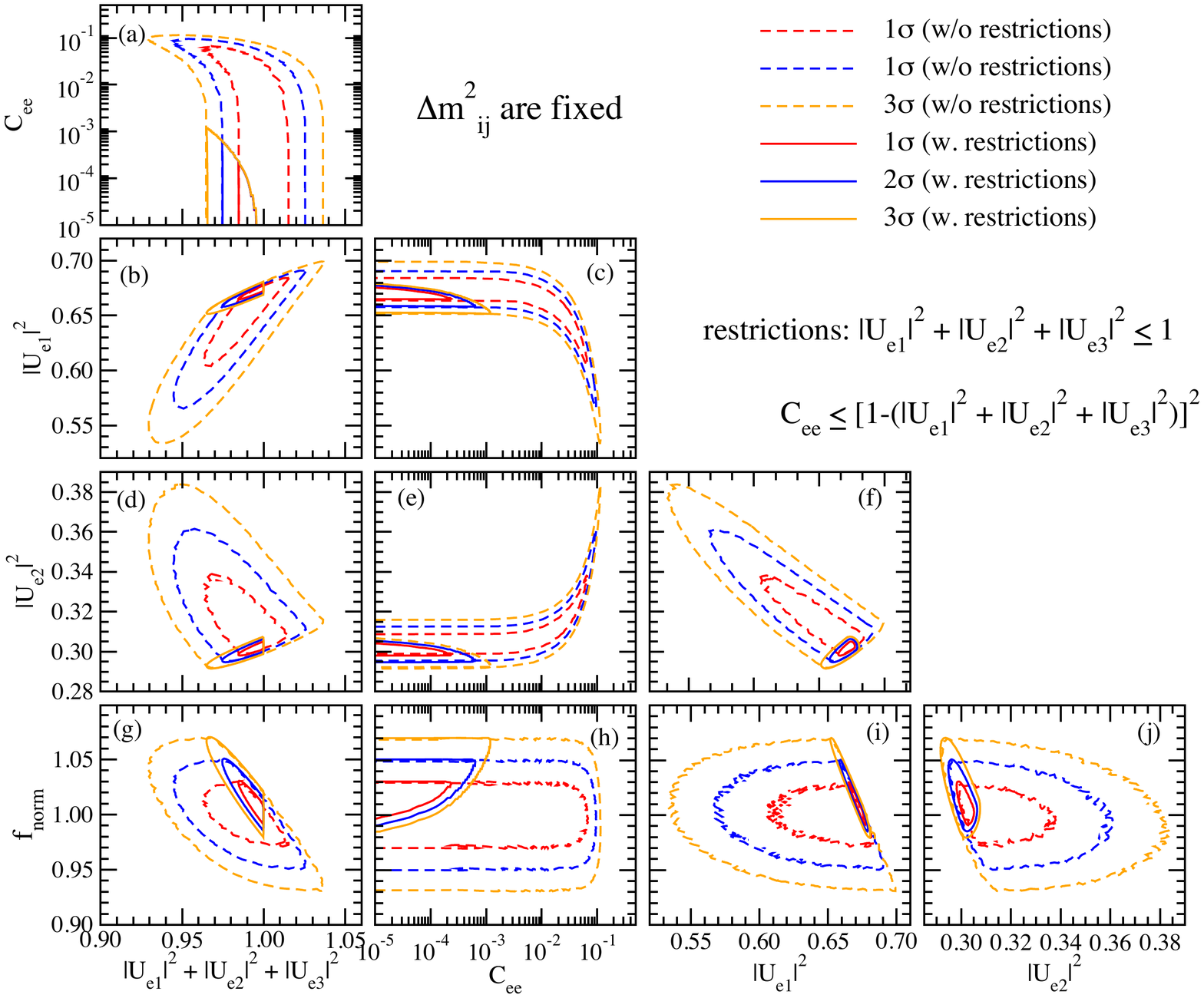}
\end{center}
\vspace{-0.9cm}
\caption{
Regions allowed at 1$\sigma$, 2$\sigma$ and 3$\sigma$ CL of five parameters $|U_{e1}|^2$, $|U_{e2}|^2$, $\sum_{i=1}^3 |U_{ei}|^2$, $\mathcal{C}_{ee}$ and $f_\text{norm}$, projected into 2 dimensional subspace. 
The case of reactor neutrino flux uncertainty of 3\%.
The solid contours are the same as shown in figure~\ref{fig:U-UV-comparison3}, that is, the cases with unitarity violation under the conditions $\sum_{i=1}^3 |U_{ei}|^2 \le 1$ and $0 \le \mathcal{C}_{ee} \le (1-\sum_{i=1}^3 |U_{ei}|^2)^2$.
Whereas the dashed contours correspond also to the cases with unitarity violation but without the above restrictions. }
\label{fig:UV-C-NC}
\end{figure}

The resultant contours of such analysis are presented in figure~\ref{fig:UV-C-NC}. The solid and the dashed contours are the cases with and without constraints (\ref{eqn:H-max-ab}). Of course, the regions outside the solid contours are unphysical in our $(3+N)$ state space unitary model. Yet, comparison between the cases with and without is revealing to understand the features of the contours, as we see below.
%

One immediately notices that there exist clear differences between 
the allowed contours obtained with and without constraints, 
both in size of the contours and the characteristic features of correlations. In particular, the features of correlations between $\vert U_{e1} \vert^2$ and $\vert U_{e2} \vert^2$ are completely different, as seen in the panel (f). That is, while $\vert U_{e1} \vert^2$ and $\vert U_{e2} \vert^2$ are positively correlated in the case with constraint, they are negatively correlated in the case without constraint with significantly prolongated contours toward left-up direction. 
Let us understand the features of the results. See section~\ref{sec:correlation-Ue12} for a model that enables us to understand the positive correlation between $\vert U_{e1} \vert^2$ and $\vert U_{e2} \vert^2$ in the case with constraint (\ref{eqn:H-max-ab}).

It appears to us that the features of the contours without constraint can be understood by the following model. Suppose that $\Delta m^2_{31}$ and $\Delta m^2_{32}$ waves cannot be discriminated by the JUNO-like setting.\footnote{
Notice that this is a completely different question from whether the JUNO setting can discriminate between $\Delta m^2_{31}$ and $\Delta m^2_{32}$ waves in the unitary case. 
}
Then, we can make approximation $|\Delta m^2_{31}| \approx |\Delta m^2_{32}| \equiv \Delta m^2_\text{atm}$, which leads to the $\bar{\nu}_e$ oscillation probability 
\begin{eqnarray} 
&& P(\bar{\nu}_e \rightarrow \bar{\nu}_e) = 
\mathcal{C}_{ee} + 
\left\{ 
\vert U_{e 1} \vert^2 + \vert U_{e 2} \vert^2 +\vert U_{e 3} \vert^2 
\right\}^2 
\nonumber \\
&-&
4 \vert U_{e 1} \vert^2 \vert U_{e 2} \vert^2 
\sin^2 \frac{ \Delta m^2_{21} x }{ 4E }
- 4 \left( \vert U_{e 1} \vert^2 + \vert U_{e 2} \vert^2 \right) 
\vert U_{e 3} \vert^2 
\sin^2 \frac{ \Delta m^2_\text{atm} x }{ 4E }. 
\label{Pee-JUNO-nores}
\end{eqnarray}

Again we can consider the effective probability by multiplying the flux normalization factor 
$f_\text{norm}$ to the probability 
as done in section~\ref{sec:correlation-Ue12}, 
$P(\bar{\nu}_e \rightarrow \bar{\nu}_e)_\text{eff} 
\equiv f_\text{norm}\times P(\bar{\nu}_e \rightarrow \bar{\nu}_e)$.
If we assume that fit to the data can separate the oscillations of the two different frequencies associated with $\Delta m^2_{21}$ and $\Delta m^2_\text{atm}$ as well as the constant term, we can obtain the following three observables, if written with the notations $x=\vert U_{e 1} \vert^2$, $y=\vert U_{e 2} \vert^2$, $z=\vert U_{e 3} \vert^2$, 
\begin{eqnarray} 
f_\text{norm}\,[\mathcal{C}_{ee} + (x+y+z)^2],  
\hspace{10mm}
f_\text{norm}\,xy, 
\hspace{10mm}
f_\text{norm}\,(x+y) z
\label{eq:x-y-z}
\end{eqnarray}
in addition to $\Delta m^2_{21}$ and $\Delta m^2_\text{atm}$. 
Clearly, the three observables cannot determine the five parameters, 
and this is the reason why the size of the allowed regions without
restrictions (dashed curves) shown in figure~\ref{fig:UV-C-NC} are much larger 
than those with restrictions (solid curves). 

Let us now focus on the problem of the correlations. 
Unlike in the case discussed in section \ref{sec:JUNO}, 
the presence of the leaking term $\mathcal{C}_{ee}$ without conditions
given in eq.~(\ref{eq:restrictions}) makes a big difference. 
First, from figure \ref{fig:UV-C-NC}(g) we notice the negative correlation 
between $f_\text{norm}$ and $x+y+z$, which is naturally 
expected to keep the energy independent term constant within some uncertainty. 
This behavior is qualitatively similar to the case with
restrictions shown  by the solid curves. 
The impact of the inclusion of $\mathcal{C}_{ee}$ without restrictions 
is to enlarge significantly the allowed region but keeping the same 
qualitative feature of anticorrelation. 

One might argue that if $x+y+z$ is decreased, $f_\text{norm}$ 
does not necessarily have to be increased since 
$\mathcal{C}_{ee}$ can be increased such that the energy independent
term (the first quantity given in (\ref{eq:x-y-z})) is kept constant, which is true. 
However, it would not be possible to keep also the coefficients of 
energy dependent terms (the second and third quantities given in (\ref{eq:x-y-z}))
simultaneously constant 
by increasing $\mathcal{C}_{ee}$ and decreasing $x+y+z$. 
Therefore, the anticorrelation between $f_\text{norm}$ and $x+y+z$ is needed
even if $\mathcal{C}_{ee}$ can be varied freely. 

However, we did not find any significant correlation between
$\mathcal{C}_{ee}$ and  $f_\text{norm}$ as we can see in figure
\ref{fig:UV-C-NC}(h).
This is because the parameter $\mathcal{C}_{ee}$ appears only once in
the probability as a constant term, completely independent from other 
terms whereas $x,y$ and $z$ appear also in the third and fourth terms. 
If $\mathcal{C}_{ee}$ is increased (decreased), $f_\text{norm}$ does not have to be 
decreased (increased) because the other parameters $x,y$ and $z$ 
can be independently adjusted (no restriction for $x,y$ and $z$) 
such that the effective probability is kept constant. 

Let us try to understand also the other correlations among 
the parameters. In particular, we focus on the ones in the panels (a), (c), (e) and (f) of figure \ref{fig:UV-C-NC}. 
For the sake of discussion let us assume that $f_\text{norm}$
is fixed to some value, e.g., to unity, as its variation seems to be not 
essential to understand the correlations we want to discuss below. 
It may be partly because, among the five parameters, $f_\text{norm}$ 
is already restricted by the pull term given in eq.~(\ref{eq:chi2_sys}). 

For a given value of $f_\text{norm}$,
if $\mathcal{C}_{ee}$ is increased, $(x+y+z)^2$ must be decreased which 
can explains the behavior we see in figure~\ref{fig:UV-C-NC} (a). 
In our analysis, the true (input) values of $x$, $y$ and $z$ (let us denote, respectively as $x_0$,
$y_0$ and $z_0$) are set to be, respectively, $x_0=$ 0.675, $y_0=$
0.303 and $z_0=$ 0.0218. 
Ignoring the small value of $z_0$, when $\mathcal{C}_{ee}$ is increased, 
$x+y$ should be decreased but keeping $xy$ constant.
We note that this is possible only if $x$ extends to the region which 
is smaller than $x_0$ and simultaneously $y$ extends to 
the region where its value is larger than $y_0$,\footnote{
One might wonder why the other possibility, increasing and decreasing, respectively,
$x$ and $y$ from $x_0$ and $y_0$, does not work. The reason is as follows.
Suppose that $x$ and $y$ are varied from $x_0$ and $y_0$ as $x_0 \to r x_0$ and
$y_0 \to y_0/r$ such that $xy$ kept constant. Then, if $\mathcal{C}_{ee}$ is increased,
we need $r x_0 + y_0/r < x_0 + y_0$, which implies $r < 1$ for $x_0>y_0$.
}
which can explain the behaviors we can see in the panels (c) and (e) of
figure \ref{fig:UV-C-NC}. By combining the results 
shown in panels (a), (c) and (e), we can also understand 
the correlations we see in the panels of (b) and (d).
Once we understand the correlations of $\mathcal{C}_{ee}-x$ (negative) 
and $\mathcal{C}_{ee}-y$ (positive), we can understand
why $x=\vert U_{e 1} \vert^2$ and $y=\vert U_{e 2} \vert^2$
are anticorrelated to each other as we can see in the panel (f) in
figure~\ref{fig:UV-C-NC}.

\section{$\tilde{ S }$ matrix elements}
\label{sec:tilde-S}

Here, we present the results of the active-active ($i j$), active-sterile ($i J$), and the sterile-sterile ($I J$) space matrix elements of $\tilde{ S }$ to first order in matter perturbation theory formulated in section~\ref{sec:UV-matter-perturbation}.
The $ii$ and $ij$ elements are
\begin{eqnarray} 
\tilde{S}_{ii}  &=& e^{ - i \Delta_{i} x} \Omega_{ii} = 
e^{ - i \Delta_{i} x} 
\left[ 
1 - i (\Delta_{A} x) \vert U_{e i } \vert^2 
+ i (\Delta_{B} x)  \sum_{\gamma} \vert U_{\gamma i } \vert^2  
\right],
\nonumber\\ 
\tilde{S}_{ij}  &=&
\Delta_{A}
U^*_{e i } U_{e j } 
\frac{ e^{ - i \Delta_{j} x} - e^{ - i \Delta_{i} x} }{ ( \Delta_{j} - \Delta_{i} )} 
- \Delta_{B} \sum_{\gamma} U^*_{\gamma i} U_{\gamma j} 
\frac{ e^{ - i \Delta_{j} x} - e^{ - i \Delta_{i} x} }{ ( \Delta_{j} -
\Delta_{i} )}
\hspace{0.5cm} (i \neq j).
\label{S-tilde-elements-3+N}
\end{eqnarray}
The $i J$ and $I j$ elements are
\begin{eqnarray} 
\tilde{S}_{i J} &=& 
\Delta_{A}
U^*_{e i} W_{e J} 
\frac{  e^{ - i \Delta_{J} x} - e^{ - i \Delta_{i} x} }
{ ( \Delta_{J} - \Delta_{i} )} 
- \Delta_{B} \sum_{\gamma} U^*_{\gamma i} W_{\gamma J} 
\frac{  e^{ - i \Delta_{J} x} - 
e^{ - i \Delta_{i} x}  }{ ( \Delta_{J} - \Delta_{i} ) },
\nonumber \\
\tilde{S}_{I j} 
&=& 
\Delta_{A} 
W^*_{e I } U_{e j} 
\frac{  e^{ - i \Delta_{j} x} - 
e^{ - i \Delta_{I} x} }{ ( \Delta_{j} - \Delta_{I} )} 
- \Delta_{B} \sum_{\gamma} W^*_{\gamma I} U_{\gamma j} 
\frac{  e^{ - i \Delta_{j} x} - e^{ - i \Delta_{I} x}  }{ 
( \Delta_{j} - \Delta_{I} ) }.
\label{S-tilde-elements-AS}
\end{eqnarray}
In sterile-sterile subspace, $\tilde{S}$ matrix is given by
\begin{eqnarray} 
\tilde{S}_{II}  &=&
e^{ - i \Delta_{I} x} \left[
1 -i ( \Delta_{A} x) \vert U_{e I} \vert^2 
+ i ( \Delta_{B} x)  \sum_{\gamma} \vert W_{\gamma I} \vert^2  
\right],
\nonumber \\
\tilde{S}_{IJ}  
&=&
\Delta_{A} 
W^*_{e I} W_{e J} 
\frac{  e^{ - i \Delta_{J} x} - e^{ - i \Delta_{I} x} }
{  ( \Delta_{J} - \Delta_{I} ) } 
- \Delta_{B} \sum_{\gamma} W^*_{\gamma I} W_{\gamma J} 
\frac{ e^{ - i \Delta_{J} x} - e^{ - i \Delta_{I} x} }{  (
\Delta_{J} - \Delta_{I} ) }
\hspace{0.2cm} (I \neq J).
\label{S-tilde-elements-SS}
\end{eqnarray}

\section{Oscillation probabilities in the disappearance channels}\label{sec:disapp-P}

The first-order matter correction in the disappearance oscillation probability $P(\nu_\alpha \rightarrow \nu_\alpha)$ is given by (next page)

\newpage
\begin{eqnarray} 
&& P(\nu_\alpha \rightarrow \nu_\alpha)^{(1)} = 
- 2 \sum_{j \neq k} 
\vert U_{\alpha j} \vert^2 
\vert U_{\alpha k} \vert^2 
\sin ( \Delta_{k} - \Delta_{j} ) x 
\left[ (\Delta_{A} x) \vert U_{e k} \vert^2 - (\Delta_{B} x) \sum_{\gamma} \vert U_{\gamma k} \vert^2  
\right] 
\nonumber\\ 
&-& 
2 \sum_{k} \sum_{J} 
\vert W_{\alpha J} \vert^2 
\vert U_{\alpha k} \vert^2 
\sin ( \Delta_{k} - \Delta_{J} ) x 
\left[ (\Delta_{A} x) \vert U_{e k} \vert^2 - (\Delta_{B} x)  \sum_{\gamma} \vert U_{\gamma k} \vert^2 
\right] 
\nonumber\\ 
&-& 
2 \sum_{j} \sum_{K} 
\vert U_{\alpha j} \vert^2 \vert W_{\alpha K} \vert^2 
\sin ( \Delta_{K} - \Delta_{j} ) x 
\left[
( \Delta_{A} x) \vert W_{e K} \vert^2 
- ( \Delta_{B} x)  \sum_{\gamma} \vert W_{\gamma K} \vert^2 
\right]
\nonumber\\  
&-& 
2 \sum_{J \neq K}  
\vert W_{\alpha J} \vert^2 \vert W_{\alpha K} \vert^2 
\sin ( \Delta_{J} - \Delta_{K} ) x 
\left[
( \Delta_{A} x) \vert W_{e K} \vert^2 
- ( \Delta_{B} x)  \sum_{\gamma} \vert W_{\gamma K} \vert^2 
\right]
\nonumber\\ 
&+& 
2 \sum_{j} \sum_{k \neq l} 
\vert U_{\alpha j} \vert^2 
\mbox{Re} \left[
\Delta_{A} 
U_{\alpha k} U^{*}_{\alpha l} 
U^*_{e k } U_{e l } 
- \Delta_{B} 
U_{\alpha k} U^{*}_{\alpha l} 
\sum_{\gamma} U^*_{\gamma k} U_{\gamma l} 
\right] 
\frac{ \cos ( \Delta_{l} - \Delta_{j} ) x - \cos ( \Delta_{k} - \Delta_{j} ) x }{ ( \Delta_{l} - \Delta_{k} )} 
\nonumber\\ 
&+& 
2 \sum_{J} \sum_{k \neq l} 
\vert W_{\alpha J} \vert^2 
\mbox{Re} \left[
\Delta_{A} 
U_{\alpha k} U^{*}_{\alpha l} 
U^*_{e k } U_{e l } 
- \Delta_{B} 
U_{\alpha k} U^{*}_{\alpha l} 
\sum_{\gamma} U^*_{\gamma k} U_{\gamma l} 
\right]  
\frac{  \cos ( \Delta_{l} - \Delta_{J} ) x - \cos ( \Delta_{k} - \Delta_{J} ) x }{ ( \Delta_{l} - \Delta_{k} )}
\nonumber\\ 
&+& 
2 \sum_{j} \sum_{l} \sum_{K} 
\vert U_{\alpha j} \vert^2 
\mbox{Re} \left[ 
\Delta_{A} 
W_{\alpha K} U^{*}_{\alpha l} 
W^*_{e K } U_{e l} 
- \Delta_{B} 
W_{\alpha K} U^{*}_{\alpha l} 
\sum_{\gamma} W^*_{\gamma K} U_{\gamma l} 
\right]
\nonumber\\ 
&\times&
\frac{ \cos ( \Delta_{l} -  \Delta_{j} ) x - \cos ( \Delta_{K} - \Delta_{j} ) x }{ ( \Delta_{l} - \Delta_{K} )} 
\nonumber\\ 
&+& 
2 \sum_{j} \sum_{k} \sum_{L} 
\vert U_{\alpha j} \vert^2 
\mbox{Re} \left[
\Delta_{A} 
U_{\alpha k} W^{*}_{\alpha L} 
U^*_{e k} W_{e L} 
- \Delta_{B} 
U_{\alpha k} W^{*}_{\alpha L} 
\sum_{\gamma} U^*_{\gamma k} W_{\gamma L} 
\right] 
\nonumber\\ 
&\times&
\frac{ \cos ( \Delta_{L} - \Delta_{j} ) x - \cos ( \Delta_{k} - \Delta_{j} ) x }{ ( \Delta_{L} - \Delta_{k} )}
\nonumber\\ 
&+& 
2 \sum_{J} \sum_{l} \sum_{K} 
\vert W_{\alpha J} \vert^2 
\mbox{Re} \left[ 
\Delta_{A} 
W_{\alpha K} U^{*}_{\alpha l} 
W^*_{e K } U_{e l} 
- \Delta_{B} 
W_{\alpha K} U^{*}_{\alpha l} 
\sum_{\gamma} W^*_{\gamma K} U_{\gamma l} 
\right]
\nonumber\\ 
&\times&
\frac{ \cos ( \Delta_{l} -  \Delta_{J} ) x - 
\cos ( \Delta_{K} - \Delta_{J} ) x }{ ( \Delta_{l} - \Delta_{K} )} 
\nonumber\\ 
&+& 
2 \sum_{J} \sum_{k} \sum_{L} 
\vert W_{\alpha J} \vert^2 
\mbox{Re} \left[
\Delta_{A} 
U_{\alpha k} W^{*}_{\alpha L} 
U^*_{e k} W_{e L} 
- \Delta_{B} 
U_{\alpha k} W^{*}_{\alpha L} 
\sum_{\gamma} U^*_{\gamma k} W_{\gamma L} 
\right]
\nonumber\\ 
&\times&
\frac{ \cos ( \Delta_{L} - \Delta_{J} ) x - \cos ( \Delta_{k} - \Delta_{J} ) x }{ ( \Delta_{L} - \Delta_{k} )} 
\nonumber\\ 
&+& 
2 \sum_{j} \sum_{K \neq L} 
\vert U_{\alpha j} \vert^2 
\mbox{Re} \left[
\Delta_{A} 
W_{\alpha K} W^{*}_{\alpha L} 
W^*_{e K} W_{e L} 
- \Delta_{B} 
W_{\alpha K} W^{*}_{\alpha L} 
\sum_{\gamma} W^*_{\gamma K} W_{\gamma L} 
\right] 
\nonumber\\ 
&\times&
\frac{ \cos ( \Delta_{L} -  \Delta_{j} ) x - \cos ( \Delta_{K} - \Delta_{j} ) x }{  ( \Delta_{L} - \Delta_{K} ) } 
\nonumber\\  
&+& 
2 \sum_{J} \sum_{K \neq L} 
\vert W_{\alpha J} \vert^2 
\mbox{Re} \left[
\Delta_{A} 
W_{\alpha K} W^{*}_{\alpha L} 
W^*_{e K} W_{e L} 
- \Delta_{B} 
W_{\alpha K} W^{*}_{\alpha L} 
\sum_{\gamma} W^*_{\gamma K} W_{\gamma L} 
\right] 
\nonumber\\ 
&\times&
\frac{ \cos ( \Delta_{L} - \Delta_{J} ) x - \cos ( \Delta_{K} - \Delta_{J} ) x }{  ( \Delta_{L} - \Delta_{K}) }.
\label{disapp-P-1st-1-6}
\end{eqnarray}

\section{Oscillation probabilities in the appearance channels}
\label{sec:app-P}

Here, we give the result of first-order matter correction term in the appearance oscillation probability $P(\nu_\alpha \rightarrow \nu_\alpha)$. For bookkeeping purpose we decompose it into the three terms: 
\begin{eqnarray} 
&& P(\nu_\beta \rightarrow \nu_\alpha)^{(1)} = 
P(\nu_\beta \rightarrow \nu_\alpha)\vert_{ \text{First} } +
P(\nu_\beta \rightarrow \nu_\alpha)\vert_{ \text{Second} } +
P(\nu_\beta \rightarrow \nu_\alpha)\vert_{ \text{Third} }.
\end{eqnarray}
The first term is given by
\begin{eqnarray} 
&& P(\nu_\beta \rightarrow \nu_\alpha)\vert_{ \text{First} }
\nonumber\\ 
&=& 
2 \sum_{j \neq k} 
\left[
- \mbox{Re} \left( U^*_{\alpha j} U_{\beta j} U_{\alpha k} U^*_{\beta k} \right)  \sin ( \Delta_{k} - \Delta_{j} ) x + \mbox{Im} \left( U^*_{\alpha j} U_{\beta j} U_{\alpha k} U^*_{\beta k} \right) \cos ( \Delta_{k} - \Delta_{j} ) x 
\right]
\nonumber\\ 
&\times&
\left[ (\Delta_{A} x) \vert U_{e k} \vert^2 - 
(\Delta_{B} x) \sum_{\gamma} \vert U_{\gamma k} \vert^2 \right] 
\nonumber\\ 
&+& 
2 \sum_{ J, k } 
\left[
- \mbox{Re} \left( W^*_{\alpha J} W_{\beta J} U_{\alpha k} U^*_{\beta k} \right)  \sin ( \Delta_{k} - \Delta_{J} ) x + \mbox{Im} \left( W^*_{\alpha J} W_{\beta J} U_{\alpha k} U^*_{\beta k} \right) \cos ( \Delta_{k} - \Delta_{J} ) x 
\right]
\nonumber\\ 
&\times&
\left[ (\Delta_{A} x) \vert U_{e k} \vert^2 - 
(\Delta_{B} x) \sum_{\gamma} \vert U_{\gamma k} \vert^2 \right] 
\nonumber\\ 
\nonumber\\ 
&+& 
2 \sum_{ j, K } 
\left[
- \mbox{Re} \left( U^*_{\alpha j} U_{\beta j} W_{\alpha K} W^*_{\beta K} \right)  \sin ( \Delta_{K} - \Delta_{j} ) x + \mbox{Im} \left( U^*_{\alpha j} U_{\beta j} W_{\alpha K} W^*_{\beta K} \right) \cos ( \Delta_{K} - \Delta_{j} ) x 
\right]
\nonumber\\ 
&\times&
\left[ (\Delta_{A} x) \vert W_{e K} \vert^2 - 
(\Delta_{B} x) \sum_{\gamma} \vert W_{\gamma K} \vert^2 \right] 
\nonumber\\ 
&+& 
2 \sum_{J \neq K} 
\left[
- \mbox{Re} \left( W^*_{\alpha J} W_{\beta J} W_{\alpha K} W^*_{\beta K} \right)  \sin ( \Delta_{K} - \Delta_{J} ) x + \mbox{Im} \left( W^*_{\alpha J} W_{\beta J} W_{\alpha K} W^*_{\beta K} \right) \cos ( \Delta_{K} - \Delta_{J} ) x 
\right]
\nonumber\\ 
&\times&
\left[ (\Delta_{A} x) \vert W_{e K} \vert^2 - 
(\Delta_{B} x) \sum_{\gamma} \vert W_{\gamma K} \vert^2 \right].
\label{app-P-1st-2nd}
\end{eqnarray}
The second term is given by
\begin{eqnarray} 
&& P(\nu_\beta \rightarrow \nu_\alpha)\vert_{ \text{Second} }
\nonumber\\ 
&=& 
2 \sum_{j} \sum_{k \neq l} 
\mbox{Re} \left[
\Delta_{A} U^{*}_{\alpha j} U_{\beta j} U_{\alpha k} U^{*}_{\beta l} U^*_{e k } U_{e l } 
- \Delta_{B} U^{*}_{\alpha j} U_{\beta j} U_{\alpha k} U^{*}_{\beta l} 
\sum_{\gamma} U^*_{\gamma k} U_{\gamma l} 
\right] 
\nonumber\\ 
&\times& 
\frac{ \cos ( \Delta_{l} - \Delta_{j} ) x - \cos ( \Delta_{k} - \Delta_{j}) x }{ ( \Delta_{l} - \Delta_{k} )} 
\nonumber\\  
&+& 
2 \sum_{j} \sum_{k \neq l} 
\mbox{Im} \left[
\Delta_{A} U^{*}_{\alpha j} U_{\beta j} U_{\alpha k} U^{*}_{\beta l} U^*_{e k } U_{e l } 
- \Delta_{B} U^{*}_{\alpha j} U_{\beta j} U_{\alpha k} U^{*}_{\beta l} 
\sum_{\gamma} U^*_{\gamma k} U_{\gamma l} 
\right] 
\nonumber\\ 
&\times& 
\frac{ \sin ( \Delta_{l} - \Delta_{j} ) x - \sin ( \Delta_{k} - \Delta_{j}) x }{ ( \Delta_{l} - \Delta_{k} )} 
\nonumber\\ 
&+& 
2 \sum_{J} \sum_{k \neq l} 
\mbox{Re} \left[
\Delta_{A} W^{*}_{\alpha J} W_{\beta J} U_{\alpha k} U^{*}_{\beta l} U^*_{e k } U_{e l } 
- \Delta_{B} W^{*}_{\alpha J} W_{\beta J} U_{\alpha k} U^{*}_{\beta l} 
\sum_{\gamma} U^*_{\gamma k} U_{\gamma l} 
\right] 
\nonumber\\ 
&\times& 
\frac{ \cos ( \Delta_{l} - \Delta_{J} ) x - \cos ( \Delta_{k} - \Delta_{J}) x }{ ( \Delta_{l} - \Delta_{k} )}
\nonumber\\ 
&+& 
2 \sum_{J} \sum_{k \neq l} 
\mbox{Im} \left[
\Delta_{A} W^{*}_{\alpha J} W_{\beta J} U_{\alpha k} U^{*}_{\beta l} U^*_{e k } U_{e l } 
- \Delta_{B} W^{*}_{\alpha J} W_{\beta J} U_{\alpha k} U^{*}_{\beta l} 
\sum_{\gamma} U^*_{\gamma k} U_{\gamma l} 
\right] 
\nonumber\\ 
&\times& 
\frac{ \sin ( \Delta_{l} - \Delta_{J}) x - \sin ( \Delta_{k} - \Delta_{J}) x }{ ( \Delta_{l} - \Delta_{k} )}
\nonumber\\ 
&+& 
2 \sum_{j} \sum_{l} \sum_{K} 
\mbox{Re} \left[
\Delta_{A} U^{*}_{\alpha j} U_{\beta j} W_{\alpha K} U^{*}_{\beta l} W^*_{e K } U_{e l } 
- \Delta_{B} U^{*}_{\alpha j} U_{\beta j} W_{\alpha K} U^{*}_{\beta l} 
\sum_{\gamma} W^*_{\gamma K} U_{\gamma l} 
\right] 
\nonumber\\ 
&\times& 
\frac{ \cos ( \Delta_{l} - \Delta_{j} ) x - \cos ( \Delta_{K} - \Delta_{j}) x }{ ( \Delta_{l} - \Delta_{K} )} 
\nonumber\\  
&+& 
2 \sum_{j} \sum_{l} \sum_{K} 
\mbox{Im} \left[
\Delta_{A} U^{*}_{\alpha j} U_{\beta j} W_{\alpha K} U^{*}_{\beta l} W^*_{e K } U_{e l } 
- \Delta_{B} U^{*}_{\alpha j} U_{\beta j} W_{\alpha K} U^{*}_{\beta l} 
\sum_{\gamma} W^*_{\gamma K} U_{\gamma l} 
\right] 
\nonumber\\ 
&\times& 
\frac{ \sin ( \Delta_{l} - \Delta_{j} ) x - \sin ( \Delta_{K} - \Delta_{j}) x }{ ( \Delta_{l} - \Delta_{K} )} 
\nonumber\\ 
&+& 
2 \sum_{J} \sum_{l} \sum_{K}
\mbox{Re} \left[
\Delta_{A} W^{*}_{\alpha J} W_{\beta J} W_{\alpha K} U^{*}_{\beta l} W^*_{e K } U_{e l } 
- \Delta_{B} W^{*}_{\alpha J} W_{\beta J} W_{\alpha K} U^{*}_{\beta l} 
\sum_{\gamma} W^*_{\gamma K} U_{\gamma l} 
\right] 
\nonumber\\ 
&\times& 
\frac{ \cos ( \Delta_{l} - \Delta_{J} ) x - \cos ( \Delta_{K} - \Delta_{J}) x }{ ( \Delta_{l} - \Delta_{K} )}
\nonumber\\ 
&+& 
2 \sum_{J} \sum_{l} \sum_{K} 
\mbox{Im} \left[
\Delta_{A} W^{*}_{\alpha J} W_{\beta J} W_{\alpha K} U^{*}_{\beta l} W^*_{e K } U_{e l } 
- \Delta_{B} W^{*}_{\alpha J} W_{\beta J} W_{\alpha K} U^{*}_{\beta l} 
\sum_{\gamma} W^*_{\gamma K} U_{\gamma l} 
\right] 
\nonumber\\ 
&\times& 
\frac{ \sin ( \Delta_{l} - \Delta_{J} ) x - \sin ( \Delta_{K} - \Delta_{J}) x }{ ( \Delta_{l} - \Delta_{K} )} .
\label{app-P-3rd-4th}
\end{eqnarray}
The third term is given by
\begin{eqnarray} 
&& P(\nu_\beta \rightarrow \nu_\alpha)\vert_{ \text{Third} }
\nonumber\\ 
&=& 
2 \sum_{j} \sum_{k} \sum_{L} 
\mbox{Re} \left[
\Delta_{A} U^{*}_{\alpha j} U_{\beta j} U_{\alpha k} W^{*}_{\beta L} U^*_{e k } W_{e L} 
- \Delta_{B} U^{*}_{\alpha j} U_{\beta j} U_{\alpha k} W^{*}_{\beta L} 
\sum_{\gamma} U^*_{\gamma k} W_{\gamma L} 
\right] 
\nonumber\\ 
&\times& 
\frac{ \cos ( \Delta_{L} - \Delta_{j} ) x - \cos ( \Delta_{k} - \Delta_{j}) x }{ ( \Delta_{L} - \Delta_{k} )} 
\nonumber\\ 
&+& 
2 \sum_{j} \sum_{k} \sum_{L} 
\mbox{Im} \left[
\Delta_{A} U^{*}_{\alpha j} U_{\beta j} U_{\alpha k} W^{*}_{\beta L} U^*_{e k } W_{e L} 
- \Delta_{B} U^{*}_{\alpha j} U_{\beta j} U_{\alpha k} W^{*}_{\beta L} 
\sum_{\gamma} U^*_{\gamma k} W_{\gamma L} 
\right] 
\nonumber\\ 
&\times& 
\frac{ \sin ( \Delta_{L} - \Delta_{j} ) x - \sin ( \Delta_{k} - \Delta_{j}) x }{ ( \Delta_{L} - \Delta_{k} )} 
\nonumber\\ 
&+& 
2 \sum_{J} \sum_{k} \sum_{L} 
\mbox{Re} \left[
\Delta_{A} W^{*}_{\alpha J} W_{\beta J} U_{\alpha k} W^{*}_{\beta L} U^*_{e k } W_{e L} 
- \Delta_{B} W^{*}_{\alpha J} W_{\beta J} U_{\alpha k} W^{*}_{\beta L} 
\sum_{\gamma} U^*_{\gamma k} W_{\gamma L} 
\right] 
\nonumber\\ 
&\times& 
\frac{ \cos ( \Delta_{L} - \Delta_{J} ) x - \cos ( \Delta_{k} - \Delta_{J}) x }{ ( \Delta_{L} - \Delta_{k} )}
\nonumber\\ 
&+& 
2 \sum_{J} \sum_{k} \sum_{L} 
\mbox{Im} \left[
\Delta_{A} W^{*}_{\alpha J} W_{\beta J} U_{\alpha k} W^{*}_{\beta L} U^*_{e k } W_{e L} 
- \Delta_{B} W^{*}_{\alpha J} W_{\beta J} U_{\alpha k} W^{*}_{\beta L} 
\sum_{\gamma} U^*_{\gamma k} W_{\gamma L} 
\right] 
\nonumber\\ 
&\times& 
\frac{ \sin ( \Delta_{L} - \Delta_{J} ) x - \sin ( \Delta_{k} - \Delta_{J}) x }{ ( \Delta_{L} - \Delta_{k} )}
\nonumber\\ 
&+& 
2 \sum_{j} \sum_{K \neq L} 
\mbox{Re} 
\left[
\Delta_{A} U^{*}_{\alpha j} U_{\beta j} W_{\alpha K} W^{*}_{\beta L} W^*_{e K} W_{e L} 
- \Delta_{B} U^{*}_{\alpha j} U_{\beta j} W_{\alpha K} W^{*}_{\beta L} 
\sum_{\gamma} W^*_{\gamma K} W_{\gamma L} 
\right] 
\nonumber\\ 
&\times& 
\frac{ \cos ( \Delta_{L} - \Delta_{j} ) x - \cos ( \Delta_{K} - \Delta_{j}) x }{ ( \Delta_{L} - \Delta_{K} )} 
\nonumber\\  
&+& 
2 \sum_{j} \sum_{K \neq L} 
\mbox{Im} 
\left[
\Delta_{A} U^{*}_{\alpha j} U_{\beta j} W_{\alpha K} W^{*}_{\beta L} W^*_{e K} W_{e L} 
- \Delta_{B} U^{*}_{\alpha j} U_{\beta j} W_{\alpha K} W^{*}_{\beta L} 
\sum_{\gamma} W^*_{\gamma K} W_{\gamma L} 
\right] 
\nonumber\\ 
&\times& 
\frac{ \sin ( \Delta_{L} - \Delta_{j} ) x - \sin ( \Delta_{K} - \Delta_{j}) x }{ ( \Delta_{L} - \Delta_{K} )} 
\nonumber\\  
&+& 
2 \sum_{J} \sum_{K \neq L} 
\mbox{Re} 
\left[
\Delta_{A} W^{*}_{\alpha J} W_{\beta J} W_{\alpha K} W^{*}_{\beta L} W^*_{e K} W_{e L} 
- \Delta_{B} W^{*}_{\alpha J} W_{\beta J} W_{\alpha K} W^{*}_{\beta L} 
\sum_{\gamma} W^*_{\gamma K} W_{\gamma L} 
\right] 
\nonumber\\ 
&\times& 
\frac{ \cos ( \Delta_{L} - \Delta_{J} ) x - \cos ( \Delta_{K} - \Delta_{J}) x }{ ( \Delta_{L} - \Delta_{K} )} 
\nonumber\\  
&+& 
2 \sum_{J} \sum_{K \neq L} 
\mbox{Im} 
\left[
\Delta_{A} W^{*}_{\alpha J} W_{\beta J} W_{\alpha K} W^{*}_{\beta L} W^*_{e K} W_{e L} 
- \Delta_{B} W^{*}_{\alpha J} W_{\beta J} W_{\alpha K} W^{*}_{\beta L} 
\sum_{\gamma} W^*_{\gamma K} W_{\gamma L} 
\right] 
\nonumber\\ 
&\times& 
\frac{ \sin ( \Delta_{L} - \Delta_{J} ) x - \sin ( \Delta_{K} - \Delta_{J}) x }{ ( \Delta_{L} - \Delta_{K} )} .
\label{app-P-5th-6th}
\end{eqnarray}

\section{Number of events for the JUNO-like setting}
\label{number-of-events}

We compute the distribution of the number of events 
coming from the inverse $\beta$-decay (IBD) reaction, 
$\bar{\nu}_e + p \to e^+ + n$, 
as a function of the visible energy by performing 
the following integral, 
\begin{eqnarray}
\frac{dN(E_{\text{vis}})}{dE_\text{vis}} 
&= & n_p t_\text{exp} \int_{m_e}^\infty {d} E_e
\int_{E_\text{min}}^\infty dE 
\sum_{i = \text{reac}}
\frac{d\phi_i(E)}{dE}
\epsilon_{\text{det}}(E_e)
\frac{d\sigma(E,E_e)}{dE_e} \nonumber \\
& & \times 
P_i(\bar{\nu}_e \to \bar{\nu}_e; L_i, E) 
R(E_e,E_{\text{vis}}),
\label{eq:event-dist}
\end{eqnarray}
where $n_p$ is the number of target (free protons), 
assumed to be $\sim 1.44 \times 10^{33}$ for 20 kt
(assuming a similar proton fraction $\simeq 12$\% as in 
the case of the Daya Bay detectors ~\cite{DayaBay:2012aa}),
$t_\text{exp}$ is the exposure, 
$\epsilon_{\text{det}}$ is the detection efficiency assumed
to be 100\% for simplicity, 
$d\phi_i(E)/dE$ is the differential fluxes of reactor neutrinos,
$d\sigma(E,E_e)/dE_e$ is the IBD cross section, 
$P_i(\bar{\nu}_e \to \bar{\nu}_e; L_i, E)$
is the $\bar{\nu}_e$ survival probabilities
for a given baseline $L_i$ and neutrino energy $E$, 
and $R(E_e,E_{\text{vis}})$ is the Gaussian resolution 
function (see below).  
We ignore the matter effect and use the probability in
vacuum as it is an excellent approximation. 
We note, however, that it is necessary to take it into account
for the precision measurement of the solar parameters, see \cite{Capozzi:2013psa,Li:2016txk}.

As a reasonable approximation for our purpose, 
we ignore the neutron recoil 
in the IBD reaction and simply assume that
neutrino energy, $E$, and the positron energy, $E_{e}$, 
is related as $E_{e} = E - (m_n-m_p) \simeq E - 1.3$ MeV. 
Due to the finite energy resolution, 
the event distribution can not be obtained 
as a function of $E_{e}$ (true positron energy) 
but as a function of the reconstructed or so called 
visible energy, $E_\text{vis}$, which is approximately related 
to neutrino energy as 
$E_\text{vis} \simeq E - (m_n-m_p) + m_e$,
after taking into account the energy resolution
(see the text below). 
Regarding the cross section, $d\sigma(E,E_e)/dE_e$, 
we use the one found in \cite{Strumia:2003zx}. 

The differential flux of reactor neutrino 
${d\phi(E)}/{dE}$ can be computed as, 
\begin{eqnarray}
\frac{d\phi(E)}{dE}   
= \frac{1}{4 \pi L^2}   
S(E) \frac{P_\text{th}}{ \langle E \rangle},
\end{eqnarray}
where $P_\text{th}$ is the thermal power of the reactor,  
${\langle E \rangle} \simeq $ 210 MeV is the average
energy released by per fission 
computed by taking into account the ratios
of the fuel compositions of the reactor (see below). 

We can replace, as a good approximation, 
the reactor complex consisting of 6 and 4 reactors, respectively, 
at Yangjiang and Taishan sites by a single reactor with 
the thermal power of 35.8 GW placed at the baseline 
$L=52.5$ km from the JUNO detector. 
We also include the contributions from 
the far reactor complexes at Daya Bay (with the baseline of 215 km)
and Huizhou (with the baseline of 265 km) sites, 
which contribute, respectively, about 3\% and 2\% 
in terms of the total number of events. 

For the reactor spectra $S(E)$, which is nothing but 
the number of neutrinos being emitted per fission per energy (MeV), 
we use the convenient analytic expressions found in ~\cite{Mueller:2011nm}
with the typical fuel compositions of the reactors, 
$^{235}$U: $^{239}$Pu: $^{238}$U: $^{241}$Pu = 0.59: 0.28: 0.07: 0.06.
For simplicity, we ignore the contributions for geoneutrinos in this
work, as it is not very important for our purpose. 

$R(E_e,E_{\text{vis}})$ is the function 
which takes into account the finite energy resolution 
of the detector and is given by 
\begin{eqnarray}
R(E_e,E_{\text{vis}})
\equiv 
\frac{1}{\sqrt{2\pi}\sigma(E_e)}
\text{exp}
\left[ -\frac{1}{2}
\left( \frac{E_e+m_e-E_{\text{vis}}}
{\sigma(E_e)}\right)^2
\right]
\label{eq:resolution-function}
\end{eqnarray}
where the energy resolution is assumed to be~\cite{Li:2013zyd}, 
\begin{eqnarray}
\frac{\sigma(E_e)}{(E_e+m_e)} = 
\frac{3\%\ }{\sqrt{(E_e+m_e)/\text{MeV}}}, 
\label{Eresolution}
\end{eqnarray}
The expected total number of events at JUNO for the 5 years of exposure 
with 100\% detection efficiency (corresponding to the total exposure of 
$\simeq3.6\times 10^3$ kt$\cdot$GW$\cdot$yr) is $\simeq1.4 \times 10^5$.

\end{document}